\documentclass[iop]{emulateapj}

\newcommand*{\teff}{$T_{\rm eff}$}
\newcommand*{\logg}{$\log~g$}
\newcommand*{\feh}{[Fe/H]}
\newcommand*{\kms}{km s$^{-1}$}
\newcommand*{\zmax}{$Z_{\rm max}$}
\newcommand*{\rmax}{$r_{\rm max}$}
\newcommand*{\rmin}{$r_{\rm min}$}

\newcommand*{\vphi}{$V_{\rm \phi}$}
\newcommand*{\vtheta}{$V_{\rm \theta}$}
\newcommand*{\vt}{$V_{\rm t}$}

\newcommand*{\rsun}{$R_\odot$}

\newcommand*{\z}{$|Z|$}
\newcommand*{\stackel}{St$\ddot{a}$ckel}

\usepackage{color}
\usepackage{epsfig}
\usepackage{graphicx}
\usepackage{natbib}
\usepackage{hyperref}

\hypersetup{
    bookmarks=true,         % show bookmarks bar?
    unicode=false,          % non-Latin characters in Acrobat??<80><99>s bookmarks
    pdftoolbar=true,        % show Acrobat??<80><99>s toolbar?
    pdfmenubar=true,        % show Acrobat??<80><99>s menu?
    pdffitwindow=true,     % window fit to page when opened
    pdfstartview={FitH},    % fits the width of the page to the window
    pdftitle={},    % title
    pdfauthor={Beers},     % author
    pdfsubject={Astronomy},   % subject of the document
    pdfcreator={dvipdf},   % creator of the document
    pdfproducer={dvipdf}, % producer of the document
    pdfkeywords={metal-poor stars},% list of keywords
    pdfnewwindow=true,      % links in new window
    colorlinks=true,       % false: boxed links; true: colored links
    linkcolor=red,          % color of internal links
    citecolor=blue,        % color of links to bibliography
    filecolor=magenta,      % color of file links
    urlcolor=cyan,           % color of external links
    breaklinks=true,
    linktocpage
}

\shorttitle{Galactic Halo Kinematics}
\shortauthors{Kim et al.}
\slugcomment{Draft version, June 13, 2019}

\begin{document}

\title{Dependence of Galactic Halo Kinematics on the Adopted Galactic Potential}

%% Use \author, \affil, plus the \and command to format author and affiliation
%% information.  If done correctly the peer review system will be able to
%% automatically put the author and affiliation information from the manuscript
%% and save the corresponding author the trouble of entering it by hand.
%%
%% The \affil should be used to document primary affiliations and the
%% \altaffil should be used for secondary affiliations, titles, or email.

%% Authors with the same affiliation can be grouped in a single
\author{Young Kwang Kim\altaffilmark{1}, Young Sun Lee\altaffilmark{1,3}, and Timothy C. Beers\altaffilmark{2}}
%        and Jinmi Yoon\altaffilmark{2}}
\altaffiltext{1}{Department of Astronomy and Space Science, Chungnam National University, Daejeon 34134, Republic of Korea}
\altaffiltext{2}{Department of Physics and JINA Center for the Evolution of the Elements, University of 
                 Notre Dame, IN 46556, USA}
\altaffiltext{3}{Corresponding author; youngsun@cnu.ac.kr}
%\altaffiltext{2}{greg.schwarz@aas.org}
%\altaffiltext{3}{AAS Journals Associate Editor-in-Chief}

\begin{abstract}

We explore differences in Galactic halo kinematic properties derived
from two commonly employed Galactic potentials: the
St$\mathrm{\ddot{a}}$ckel potential and the default Milky Way-like
potential used in the ``Galpy'' package (MWPotential2014), making use of
stars with available metallicities, radial velocities, and proper
motions from Sloan Digital Sky Survey Data Release 12. Adopting the
\stackel\ potential, we find that the shape of the metallicity
distribution function (MDF) and the distribution of orbital rotation
abruptly change at \zmax\ = 15 kpc and \rmax\ = 30 kpc (where \zmax\ and
\rmax\ are the maximum distances reached by a stellar orbit from the
Galactic plane and from the Galactic center, respectively), indicating
that the transition from dominance by the inner-halo stellar population
to the outer-halo population occurs at those distances. Stars with
\zmax\ $>$ 15 kpc show an average retrograde motion of \vphi\ =
--60 \kms, while stars with \rmax\ $>$ 30 kpc exhibit an even larger
retrograde value, \vphi\ = --150 \kms. This retrograde signal 
is also confirmed using the sample of stars with radial velocities
obtained by $Gaia$ Data Release 2, assuming the \stackel\ potential. In comparison, when using the
shallower Galpy potential, a noticeable change in the
MDF occurs only at \zmax\ = 25 kpc, and a much less extreme retrograde
motion is derived. This difference arises because stars with highly retrograde
motions in the \stackel\ potential are unbound in the shallower Galpy
potential, and stars with lower rotation velocities reach larger \zmax\
and \rmax. The different kinematic characteristics derived from the two
potentials suggest that the nature of the adopted Galactic potential can
strongly influence interpretation of the properties of the Galactic
halo.

\end{abstract}


\section{Introduction} \label{sec:intro}

Although comprising only about 1\% of the Galaxy's total stellar mass,
the stellar halo of the Milky Way (MW) provides valuable clues to its
formation and evolutionary history. Recent studies (e.g.,
\citealt{yanny2000,newberg2002,majewski2003,belokurov2006, bonaca2012})
from large-scale sky surveys such as the Sloan Digital Sky Survey (SDSS;
\citealt{york2000}) have shown that the MW's halo includes multiple 
stellar components and continually accretes stars from dwarf galaxies 
disrupted by the MW's tidal forces. Thanks to the long dynamical 
timescale for accreted stars to be completely mixed into the halo, 
the Galactic halo serves as a fossil record, revealing its 
accretion history (\citealt{Bland-Hawthorn2016}).

The $\Lambda$-cold dark matter ($\Lambda$-CDM) scenario predicts
that large galaxies, such as the MW, formed via hierarchical mergers
(\citealt{white1991}). The majority of the halo is built at early
times, with most of the halo stars coming from the mergers of a few massive
satellites (\citealt{bullock2005}). According to more
sophisticated simulations of galaxy formation, the halos of galaxies
like the MW have a dual origin, and the inner and outer regions of their
halos might be dominated by accreted stellar populations with
different chemical and kinematic properties, respectively
(\citealt{abadi2006, zolotov2009, font2011, mccarthy2012, tissera2012,
tissera2013, tissera2014, cooper2015}).

From a kinematic analysis of halo stars within 4 kpc of the Sun,
\citet{carollo2007} first argued that the Galactic halo is composed of
at least two distinct stellar populations --- an inner-halo population
(IHP) and an outer-halo population (OHP), separable by their different
spatial distribution, metallicity, and kinematics. The inner-halo
component dominates the population of stars found at distances up to 10
-- 15 kpc from the Galactic center, while the outer-halo component is
dominant in the region beyond 15 -- 20 kpc. The inner halo also exhibits
a flatter density profile than the nearly spherical outer halo. The
metallicity distribution function (MDF) of the inner-halo stars peaks at
[Fe/H] = --1.6, while the peak of the MDF for the outer-halo stars occurs 
at [Fe/H] = --2.2. Kinematically, the IHP shows either zero or slightly
prograde rotation and eccentric orbits, while the OHP exhibits a net
retrograde rotation of about --80 \kms\ on more circular (or less
eccentric) orbits (\citealt{carollo2007,carollo2010,beers2012}).

However, \citet{schonrich2011, schonrich2014} claimed that the results by
\citet{carollo2010} might have arisen due to faulty distance
determinations and unaccounted-for selection biases; their reanalysis of
the data used by \citet{carollo2010} does not support the claim of
retrograde motion for the OHP. \citet{schonrich2014} further argued that
the results of \citet{carollo2007, carollo2010} needed to account for a
possible metallicity bias in their selection and verify
whether or not the claimed retrograde motion is due to observational
errors. In the meantime, \citet{beers2012} disputed these
arguments and reported that the result of \citet{schonrich2011} is due,
at least in part, to their adoption of an incorrect main-sequence
absolute-magnitude relationship. Furthermore, they demonstrated
that the retrograde signature of the OHP appears using proper motions
alone. This interpretation is also supported by a recent study (\citealt{tian2019}) of local
K-giant stars selected from the Large Sky Area Multi-Object Fiber Spectroscopic Telescope
(LAMOST; \citealt{cui2012}). In addition, many recent
studies of in situ halo stars up to tens of kpc away from the Galactic
center consistently support the dual nature of the Galactic halo \textcolor{red}{(e.g.,
\citealt{dejong2010,deason2011,an2013,an2015,kafle2013,
kafle2017,allendeprieto2014,chen2014,fernandez-alvar2015,das2016})}.

For in situ studies of the most distant halo kinematics, one can only
rely on radial-velocity information (e.g., \citealt{deason2011,
kafle2013,kafle2017}). However, it is feasible to consider the full
six-dimensional location and velocity phase space of halo stars reaching
large Galactocentric distance derived from stars in the Solar
neighborhood, where proper motions can be readily measured. In this
case, however, a Galactic gravitational potential must be adopted in
order to compute stellar orbital parameters, such as eccentricity, apo-Galacticon and
peri-Galacticon distances, and maximum distance from the Galactic midplane,
which are used to separate the stellar populations in
the Galactic halo system. Even though it is recognized that the adoption of
different potentials (and masses) can influence the interpretation of
the chemical and kinematic properties of the halo, to date there has not
been $quantitative$ investigation of these effects. This is the key
motivation for pursuing the present study.

In this paper, we consider two well-known and often-adopted Galactic
potentials---a \stackel-type potential (hereafter referred to as ``the
\stackel\ potential'') employed by  \citet{chiba2000} and
\citet{carollo2007,carollo2010}, and the \textrm{MWPotential2014} 
(hereafter referred to as ``the Galpy potential'') included in the Galpy
package by \citet{bovy2015}---and examine the resulting influence on the
derived chemical and kinematic properties, as well as the corresponding
interpretation of the nature of the MW halo system. 

%\textbf{Note that 
%we do not attempt to prove whether or not the dual nature of the 
%Galactic halo exists, but just compare the orbital parameters 
%under the two potentials to check how the inferred property of the Galactic 
%halo could differ between them.}

The major difference between the two potentials 
is the assumed total mass of the MW. The total mass adopted in the
\stackel\ potential exceeds that of the Galpy potential; hence the
\stackel\ potential well is deeper than that of the Galpy potential.
This results in larger numbers of stars inferred to be bound
(total energy (TE) $<$ 0) to the Galaxy in the \stackel\ potential compared
with the Galpy potential. Accordingly, derived apo-Galacticon distances
and maximum distances from the Galactic midplane for commonly bound stars 
in both potentials are larger in the Galpy potential than in the \stackel\ potential.  

This paper is outlined as follows. We describe the criteria used to
select a local sample for our study in Section~\ref{sec:sel}.
Section~\ref{sec:orbit} addresses the derivation of velocity components
and orbital parameters for the stars in our sample. In
Section~\ref{sec:results}, we compare the chemical and kinematic
properties of the Galactic halo based on orbital parameters derived from
both potentials. We examine the impact of distance errors and
target selection effects on the chemical and kinematic properties of the
Galactic halo in Section \ref{sec:bias}. Section~\ref{sec:discussion}
discusses our findings on the nature of the Galactic halo, and a summary
follows in Section \ref{sec:summary}.

\section{Selection of Local Sample Stars} \label{sec:sel}

The sample used in this study consists of spectrophotometric and
telluric calibration stars observed during the course of the legacy
SDSS and Sloan Extension for Galactic Exploration and Understanding
(SEGUE; \citealt{yanny2009}) as well as the Baryon Oscillation Spectroscopic
Survey (BOSS; \citealt{daw2013}). The apparent magnitude and 
color ranges of the calibration stars are $15.5 < g_0 < 18.5$, and 
$0.6 < (u-g)_0 < 1.2$ and $0.0 < (g-r)_0 < 0.6$, after correcting for the effects 
of interstellar absorption and reddening based on the dust map of 
\citet{schlegel1998}. The calibration stars are assembled from the SDSS Data Release 12 
(SDSS DR12; \citealt{alam2015}); the total number of stars is $55,293$.  

Stellar atmospheric parameters (\teff, \logg, and [Fe/H]) for the 
calibration stars were estimated using the SEGUE Stellar Parameter Pipeline
(SSPP; \citealt{allendeprieto2008, lee2008a,lee2008b,lee2011, smo2011})
from the medium-resolution ($R \sim 2000$) SDSS spectra. Radial
velocities were adopted from the SDSS pipeline; they have precisions of
5 -- 20 \kms, depending on the signal-to-noise ratio (S/N) of the spectrum, 
and negligible zero-point errors \citep{yanny2009}. For stellar
distance estimates, we employed the methods of \citet{beers2000,
beers2012}; the reported uncertainty is on the order of 15--20\%.
Proper motions, which are corrected for known systematic errors, 
were obtained from the SDSS DR12 (\citealt{munn2004,munn2008}). 
Even though the proper motions from the $Gaia$
Data Release 2 ($Gaia$ DR2; \citealt{gaia2018}) are now available, 
we used the proper motions from the SDSS DR12 in order to
enable a more direct comparison with previous studies (e.g.,
\citealt{carollo2010}). 
We adopted the criteria used by \citet{carollo2010} to select the final
sample for our analysis. Those criteria are as follows:

\begin{itemize}
\item The spectra must have an average S/N greater
than 10/1, as well as effective temperatures in the range 4500 $\leq$ \teff\
$\leq$ 7000~K.

\item Stars must have a measured radial velocity with reported error better than
20 \kms, as well as proper motions with errors less than 4 mas yr$^{-1}$
(or relative proper motion errors less than 10\%).

\item Stars must have distances of $d < 4$ kpc from the Sun.

\item Stars must reside in the distance range of $6 < R < 10$ kpc from the Galactic center, 
projected onto the Galactic plane.
\end{itemize}

In our study, we considered two samples. The first sample, which 
we refer to as the ``DR7 sample,'' consists of 18,821 
calibration stars of the SDSS Data Release 7 (SDSS DR7;
\citealt{abazajian2009}), subselected from the SDSS DR12. This is the same sample used by \citet{carollo2010}. 
The other one is an extended sample of 29,447 calibration stars from the SDSS DR12, 
which we refer to as the ``DR12 sample.'' The DR12 sample also includes
stars from the DR7 sample. The reason for using the same sample as
Carollo et al. is that we want to check whether we are able to derive the
same Galactic halo properties from an application of a different
Galactic potential. The much larger DR12 sample is used to check whether we
obtain evidence for the dual nature of the Galactic halo from the two different
Galactic potentials considered.

\section{Calculations of Space Velocity Components and Orbital Parameters} \label{sec:orbit}

Based on the distances, radial velocities, and proper motions adopted
for our sample of stars, we derived their space velocity components in a
spherical coordinate system, after correcting for the solar motion. We
adopted $V_{\rm LSR}$ = 220 \kms\ for the rotation motion of the local
standard of rest (LSR) and the Sun's position of \rsun\ = 8 kpc from the
Galactic center (note that \citet{carollo2010} used 8.5 kpc). For the
solar peculiar motion with respect to the LSR, we assumed ($U$, $V$,
$W$)$_{\odot}$ = (--10.1, 4.0, 6.7) \kms\ (\citealt{hogg2005}), where
the velocity components $U$, $V$, and $W$ are positive in the direction
toward the Galactic anticenter, Galactic rotation, and north Galactic
pole, respectively. In our adopted system, a disk star has a prograde
rotation of \vphi\ = 220 \kms\ (\citealt{kerr1986}); the retrograde rotation
is indicated by \vphi\ $ < 0$ \kms. Stars with $V_{\rm r} > 0$ \kms\
move away from the Galactic center, and stars with \vtheta\ $> 0$ \kms\
move toward the south Galactic pole. We also define the tangential
velocity, \vt\ = $\sqrt{{V_{\rm \theta}}^2 + {V_{\rm \phi}}^2}$, and the
angular momentum, $|\vec{L}| = rV_{\rm t}$, where $r$ is the distance
from the Galactic center to a given star.

We also make use of a Galactocentric Cartesian reference frame
denoted by $(X,Y,Z)$, where the axes are positive in orientation toward
the Sun, Galactic rotation, and north Galactic pole, respectively.
We introduce an angle ($\alpha$) between the
orientation of the angular momentum vector and the negative $Z$ axis, defined
by:

\begin{eqnarray*}
\mathrm{ {\alpha} = {\cos}^{-1} \Big( \frac{l_Z}{|\vec{L}|} \Big)
= {\cos}^{-1} \Big( \frac{RV_{\phi}}{rV_t} \Big) } \\
\mathrm{ = {\cos}^{-1} \bigg\{ \Big( \frac{R}{r} \Big) \frac{(\frac{V_{\phi}}{|V_{\phi}|})} {\sqrt{1+ (\frac{V_{\theta}}{|V_{\phi}|})^2} } \bigg\} },
\end{eqnarray*}
%\begin{equation}
%{\alpha} = {\cos}^{-1} \Big(\frac{l_Z}{|\vec{L}|}\Big)
%= {\cos}^{-1} \Big(\frac{RV_{\phi}}{rV_t}\Big) \\
%~~= {\cos}^{-1} \bigg\{\Big(\frac{R}{r}\Big) \frac{(\frac{V_{\phi}}{|V_{\phi}|})}{\sqrt{1+(\frac{V_{\theta}}{|V_{\phi}|}%)^2}\bigg\}},
%\end{equation}

\noindent where $l_{\rm Z}$ is the $Z$ component of the angular momentum
and positive in the negative $Z$-axis direction. In this notation, stars
with $l_{\rm Z} > 0$ or \vphi\ $>0$ have prograde orbits and $\alpha <
90^{\circ}$. Retrograde orbits have $\alpha > 90^{\circ}$ and the
inclination ($i$) of their orbital plane increases as
$\mathrm{{\alpha}}$ approaches $90^{\circ}$. For prograde motions, the
inclination angle $i$ = $\alpha$; for retrograde orbits, $i$ =
$180^{\circ}-{\alpha}$. 

We used both the \stackel\ and the Galpy gravitational potentials to
calculate the orbital parameters of our program stars. The \stackel\
potential is analytic, consisting of a flattened, oblate disk and a
nearly spherical massive dark matter halo. 
The tidal cutoff radius is 200 kpc, and the disk mass is 
$M_{\rm d} = 9.0 \times 10^{10}~M_{\odot}$. The central density and mass
of the halo are $\rho_0 = 2.45 \times 10^7 ~M_{\odot}~\rm{kpc}^{-3}$ and
$M_{\rm h}~(r<200~ \rm{kpc}) = 2.2 \times 10^{12}~M_{\odot}$,
respectively (\citealt{dezeeuw1986,dejonghe1988,sommer1990,chiba2000}).

The Galpy potential comprises three components. The bulge
is parameterized as a power-law density profile that is exponentially
cut off at 1.9 kpc with a power-law exponent of --1.8. The disk is
represented by a Miyamoto--Nagai potential with a radial scale length of
3 kpc and a vertical scale height 280 pc. The halo is modeled as a
Navarro--Frenk--White halo with a scale length of 16 kpc. The bulge and disk 
masses of the Galpy potential are $M_{\rm b} = 0.5 \times 10^{10}~M_{\odot}$ and 
$M_{\rm d} = 6.8 \times 10^{10}~M_{\odot}$, respectively. The virial mass 
of the Galpy potential is $M_{\rm{vir}} = 0.8 \times 10^{12}~M_{\odot}$ 
(\citealt{bovy2015}). 

The orbital parameters of our sample of stars (for each potential), 
such as the perigalactic distance (\rmin, the minimum distance of
an orbit from the Galactic center), and the apogalactic
distance (\rmax, the maximum distance of an orbit from the Galactic center) 
as well as \zmax\ (the maximum distance of a stellar orbit above
or below the Galactic plane), are derived from an analytic solution 
for the \stackel\ potential (see the equations of motion in
\citet{sommer1990}), and integrating their orbital paths for a time of 11 Gyr using the 
Galpy package (\citealt{bovy2015}) for the Galpy potential, based on 
the starting position and observed velocities. We describe the method 
for estimating the uncertainties in the derived parameters in 
Section \ref{sec:error}.

These orbital calculations allowed us to obtain, for the DR7 sample orbital parameter, 
estimates for a total number of $N = 18,749$ and $18,370$ stars for the \stackel\ and 
Galpy potentials, respectively. If a star is
unbound in the \stackel\ potential, we eliminated it from further
consideration, so by definition all stars in this potential are bound.
The shallower Galpy potential has additional numbers of unbound stars.
The total number of stars for the DR12 sample with orbital
parameter estimates is 29,273 for the \stackel\ potential and 28,483 for
the Galpy potential. Note that when we mention unbound stars throughout
the rest of the paper, we always mean those that are unbound under the
Galpy potential.

\section{Results} \label{sec:results}

For simplicity, we focus on the distributions of rotation velocity
(\vphi) and the MDF for comparison of the derived halo properties for
the two potentials in our analysis. This also allows for a straightforward
comparison with previous studies based on local samples of halo stars,
as described below.

\subsection{Differences in Halo Properties from the DR7 Sample} \label{sec:DR7}

% FIGURE 1
\begin{figure*}[t]
%\centering
%\figurenum{1}
\epsscale{0.9}
\plotone{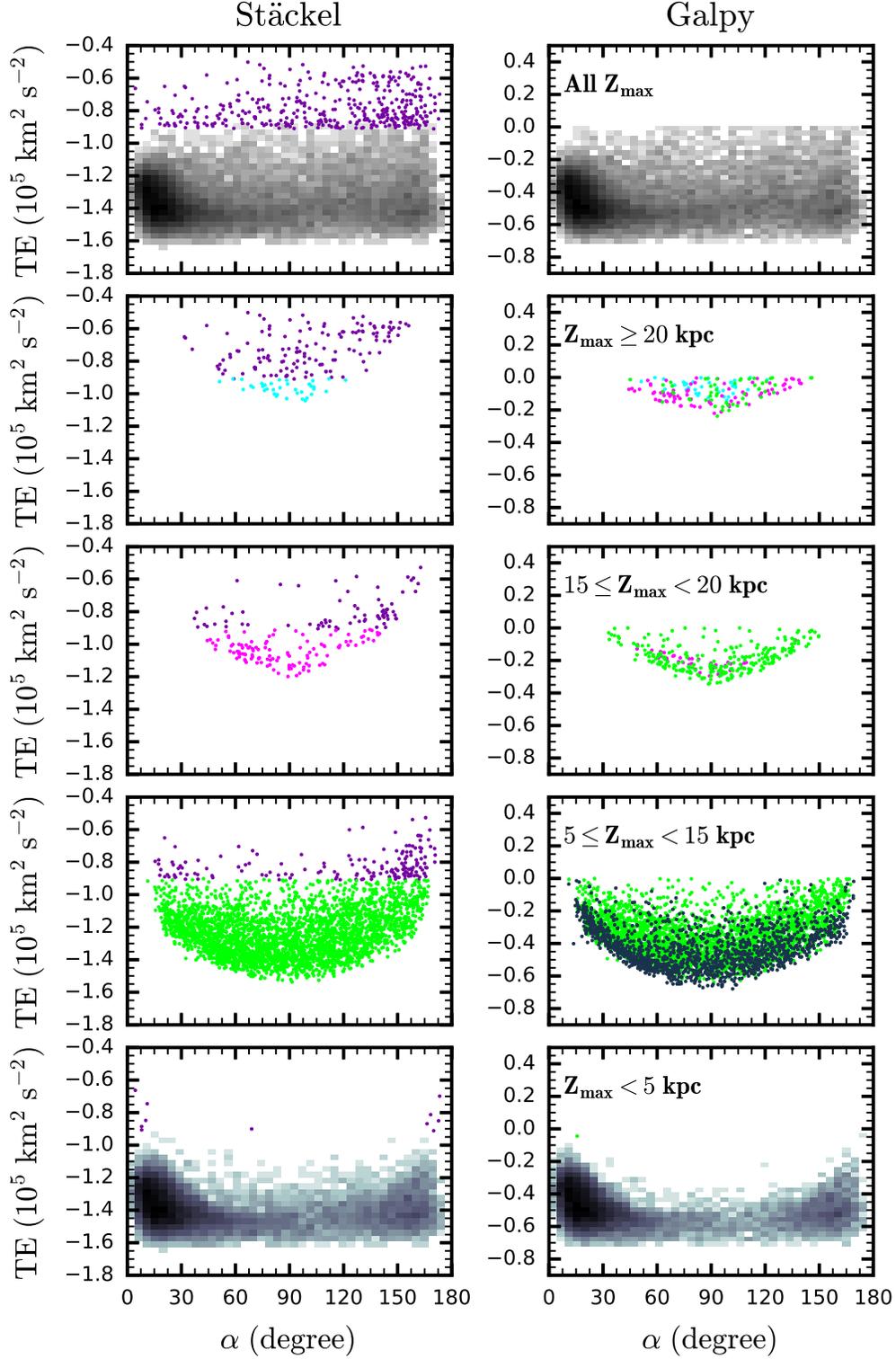}
\caption{Distribution of total energy (TE) versus $\alpha$ for the \stackel\ (left panels)
and the Galpy (right panels) potentials, respectively. Note that the scale
of the total energy differs between the two potentials in different
ranges of \zmax, as listed in the right panels. Here, $\alpha$ is the angle
between the direction of the angular momentum vector and the negative
$Z$-axis for each star. Maps in the top and bottom panels display the
number density (low to high from bright to dark) on a log-10 based scale
for all bound stars and stars with \zmax\ $<$ 5 kpc, respectively. In
the left panels, stars in different colors represent objects in
respective \zmax\ cuts from the \stackel\ potential. Stars in purple are
bound objects in the \stackel\ potential, but unbound objects in the
Galpy potential. The right panels also show the stars in respective
\zmax\ cuts from the Galpy. The green points from the first to the
fourth from the bottom in the right column of panels are for stars in the
range of 5 $\leq$ \zmax\ $<$ 15 kpc in the
\stackel\ potential. The black dots in the second panel 
from the bottom are the stars with \zmax\ $<$ 5 kpc
in the \stackel\ potential. Similarly, the magenta stars in the right 
panels are in the range of 15 $\leq$ \zmax\ $<$ 20 kpc in the \stackel\ 
potential, while the cyan stars in the right panels are in the range 
of \zmax\ $\geq$ 20 kpc in the \stackel\ potential.}
\label{figure1}
\end{figure*}

% FIGURE 2
\begin{figure*}[t]
%\centering
%\figurenum{1}
\epsscale{1.15}
\plotone{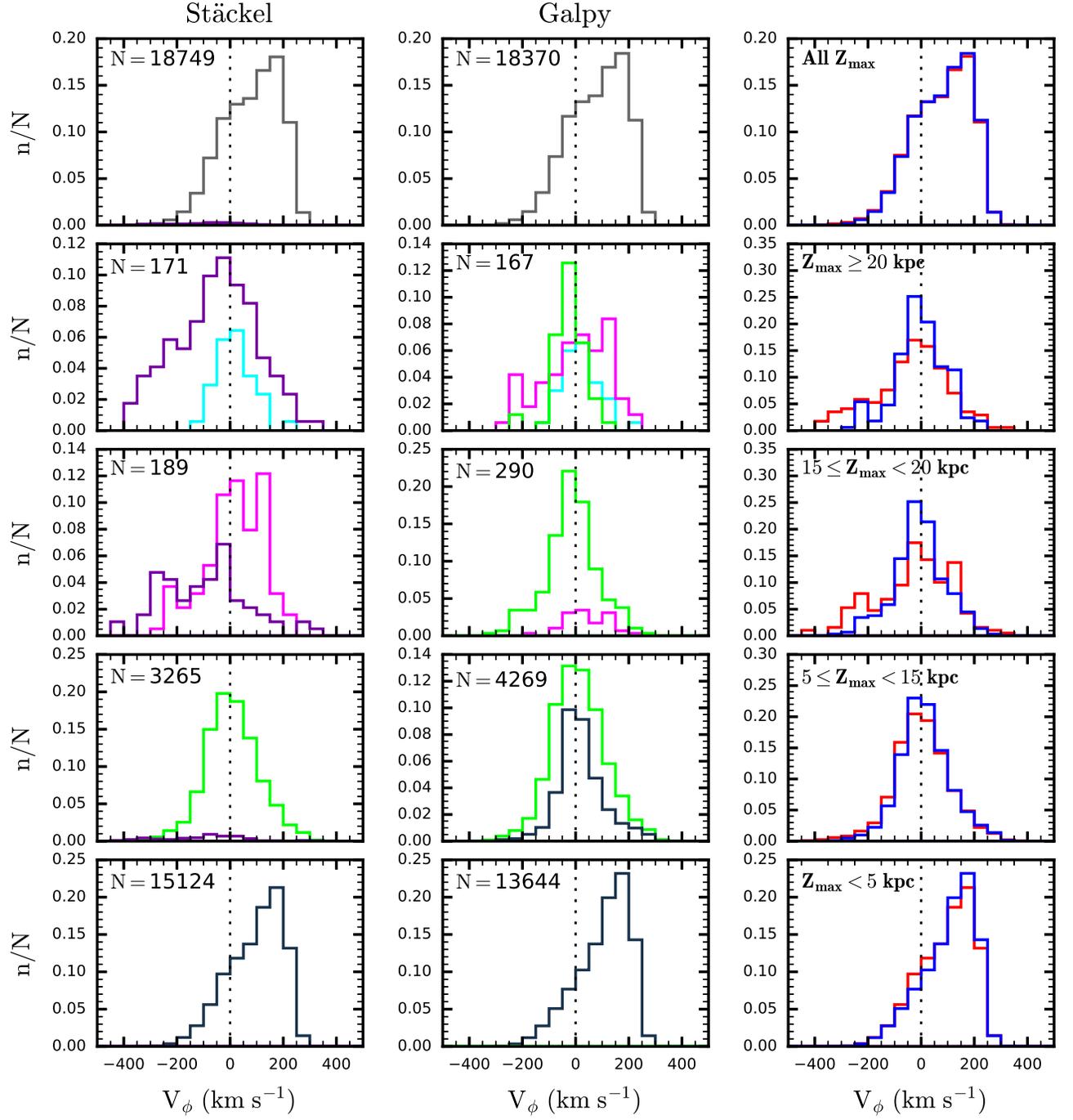}
\caption{Fractional distribution of the rotation velocities in different bins of
\zmax, as listed in the right panels, for the \stackel\ (left panels) and
the Galpy (middle panels) potentials. Here, $N$ is the total 
number of stars in each bin. The definition for each color is the same as in Figure \ref{figure1}.
The right-column panels display the distribution of rotation velocity for
all bound stars in given bin of \zmax\ for the \stackel\ potential as
red lines, and for the Galpy potential as 
blue lines.}
\label{figure2}
\end{figure*}

% FIGURE 3
\begin{figure*}
\begin{center}
%\figurenum{2}
\epsscale{1.15}
\plotone{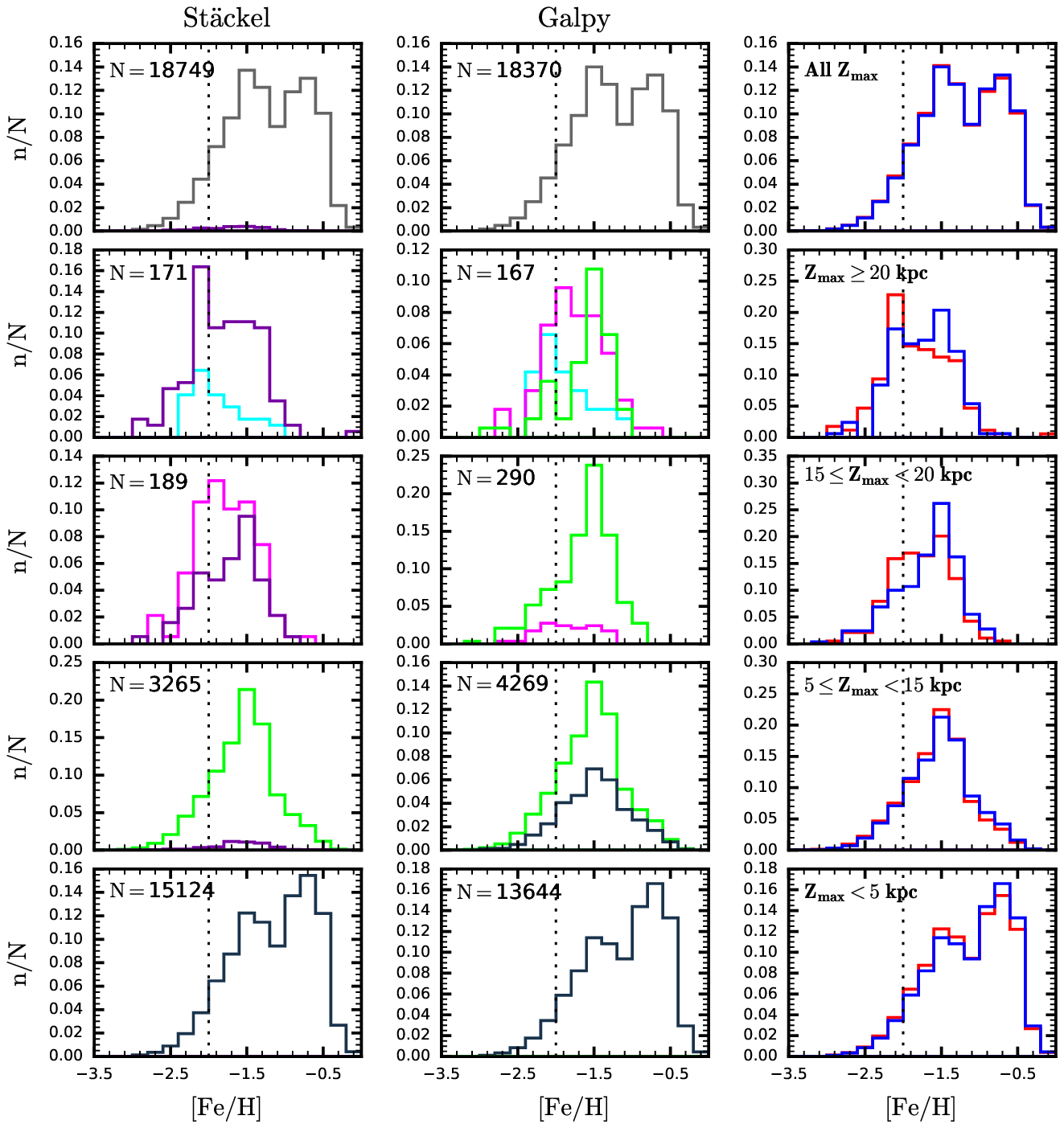}
\caption{Same as in Figure \ref{figure2}, but for the metallicity distribution functions. 
The legend for each color is the same as in Figure \ref{figure1}. The vertical dotted line 
is at [Fe/H] = --2.0 for reference.}
\label{figure3}
\end{center}
\end{figure*}

In order to compare the derived halo properties between the \stackel\
and Galpy potential, we examined the distribution of TE
as a function of $\alpha$, as shown in Figure~\ref{figure1}. As
described in the previous section, $\alpha$ is the angle between the
angular momentum vector and the direction of the negative $Z$-axis.
 
The left- and right-column panels in the figure show the stars in 
the \zmax\ bin listed in the right-column panels, based on the \stackel\
and the Galpy potential, respectively. The gray (top panels) and black
(bottom panels) maps exhibit the logarithmic number density (low to high
from bright to dark) for all bound stars and stars in the range \zmax\
$<$ 5 kpc, respectively. In the left column of panels, stars with
different colors represent objects in respective \zmax\ cuts from the
\stackel\ potential. Stars in purple are bound in the \stackel\
potential, but unbound in the Galpy potential. In the middle three rows
of panels, stars in black with \zmax\ $<$ 5 kpc in the \stackel\
potential fall in the ranges \zmax\ $<$ 5 kpc and 5 $\leq$ \zmax\ $<$ 15
kpc in the Galpy potential, while stars in green with 5 $\leq$ \zmax\
$<$ 15 kpc of the \stackel\ potential are located in all \zmax\ cuts of
the Galpy potential. Stars in magenta with 15 $\leq$ \zmax\ $<$ 20 kpc
in the \stackel\ potential occupy the ranges of 15 $\leq$ \zmax\ $<$ 20
kpc and \zmax\ $\geq$ 20 kpc of the Galpy potential. Stars with cyan
colors with \zmax\ $\geq$ 20 kpc in the \stackel\ potential fall in the
range \zmax\ $\geq$ 20 kpc in the Galpy potential as well.

Inspection of Figure~\ref{figure1} reveals the following. As \zmax\
increases, most of the stars tend to cluster around $\alpha=90^{\circ}$.
For a given interval of \zmax, the stars that are closer to
$\alpha=90^{\circ}$ have the lowest TE, while the stars progressively
farther from $\alpha=90^{\circ}$ have larger TE. Considering only the
bound stars from each potential, the stars with \zmax\ $>$ 5 kpc show a
rather symmetric distribution with respect to $\alpha=90^{\circ}$. The
green stars reaching high \zmax\ in the shallower Galpy
potential have lower \zmax\ in the \stackel\ potential, but
are also symmetric around $\alpha=90^{\circ}$. In particular, we notice
that some portion of stars (green dots in the second from the top in the
right panels) in the range $5 \leq$ \zmax\ $<15$ kpc in the \stackel\
potential reach the region of \zmax\ $>$ 20 kpc in the Galpy potential.
The unbound stars (shown in purple in the left panels) are more
populated in the region of $\alpha > 90^{\circ}$.

% FIGURE 4
\begin{figure*}[!t]
\plotone{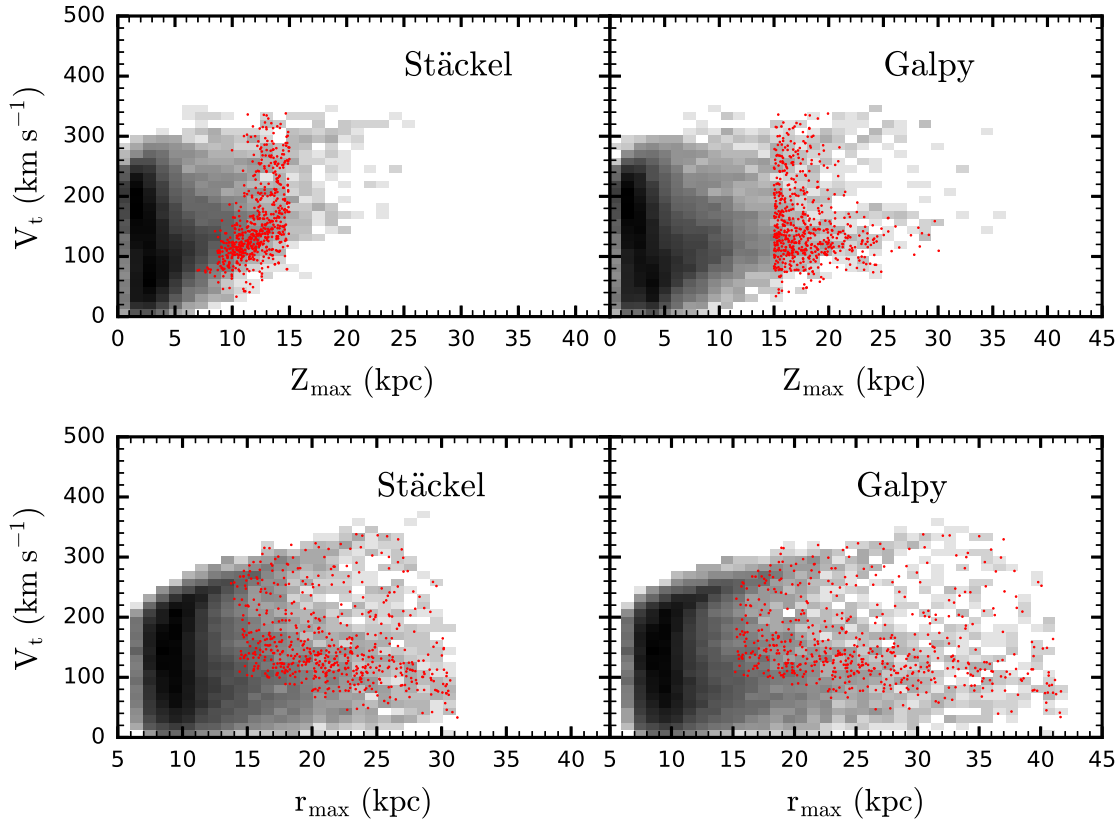}
\caption{Distribution of tangential velocity (\vt) as a function of 
\zmax\ (top panels) and \rmax\ (bottom panels) for the \stackel\ (left panels) 
and the Galpy (right panels) potential, respectively. In these plots, bound stars 
are displayed with the number density map on a log-10 based scale (low to high 
from bright to dark), whereas red dots mark stars with \zmax\ $<$ 15 kpc
from the \stackel\ potential and with \zmax\ $>$ 15 kpc from the Galpy
potential.}
\label{figure4}
\end{figure*}

Figure~\ref{figure2} shows histograms of the rotation velocities in
different bins of \zmax\ for the \stackel\ (left panels) and the Galpy
(middle panels) potential, respectively. The right column of panels is
the distribution of rotation velocity for all stars in a given bin of
\zmax\ for the \stackel\ (red line) and the Galpy potential (blue line).
The general trend in this figure, except for the region (\zmax\ $<$ 5
kpc) in which thick-disk stars dominate, is that the distribution of
rotation velocities of bound stars (the green, magenta, blue, and black
histograms) is symmetric around \vphi\ = 0, and the net rotational
velocity is nearly zero in both potentials. On the other hand, the
rotational motion of the unbound stars (purple histogram in the left
panels) in the Galpy potential above \zmax\ = 5 kpc is predominantly
retrograde, regardless of the \zmax\ range. As can be seen in the
right-column panels, the result is that the \vphi\ distribution (red
histogram) of the bound stars in the \stackel\ potential exhibits a
retrograde motion (blue histogram) over \zmax\ $>$ 15 kpc, while the net
rotation velocity of bound stars in the Galpy potential is nearly zero.
This leads us to conclude that the pronounced differences in the
\vphi\ distribution over \zmax\ between the two potentials are primarily 
caused by the absence of the unbound stars in the Galpy potential.

The MDFs are shown in Figure \ref{figure3}, in different bins of
\zmax\ derived from the \stackel\ (left panels) and the Galpy (middle
panels) potentials, respectively. The right-column panels are the MDFs
for the bound stars in each range of \zmax\ for the \stackel\ (red line)
and the Galpy (blue line) potential, respectively. In the left-column
panels of the figure, we note that the MDF for bound stars in the
\stackel\ potential exhibits a transition from relatively metal-rich
(--1.7 $\leq$ [Fe/H] $<$ --1.0) to more metal-poor ([Fe/H] $<$ --1.7)
beyond \zmax\ = 15 kpc. In contrast, the unbound stars (purple
distribution) in the Galpy potential exhibit a peak of the MDF in the
metal-poor region over \zmax\ = 20 kpc, although there is still some
fraction of stars in the metal-rich region. 

For the Galpy potential, the shape of the MDF does not change much
between \zmax\ = 5 and 20 kpc. However, beginning at \zmax\ $>$ 20 kpc,
the MDF exhibits a larger fraction of more metal-poor stars, with
substantial numbers of relatively high-metallicity ([Fe/H] $>$ --1.7)
stars, displaying two apparent peaks. Interestingly, the metal-rich stars are
contributed from the stars in the range 5 $\leq$ \zmax\ $<$ 15 kpc in
the \stackel\ potential. Consequently, the metallicity shift occurs at
larger \zmax\ ($>$ 20 kpc) for the Galpy potential than for the
\stackel\ potential. The difference in the \vphi\ distribution and MDF
over \zmax\ between the two potentials suggests that the adoption 
of the Galpy potential may result in a different interpretation of the Galactic
halo properties, compared to that from obtained with the
\stackel\ potential.

% FIGURE 5
\begin{figure*}[t]
%\figurenum{4}
\epsscale{1.0}
\plotone{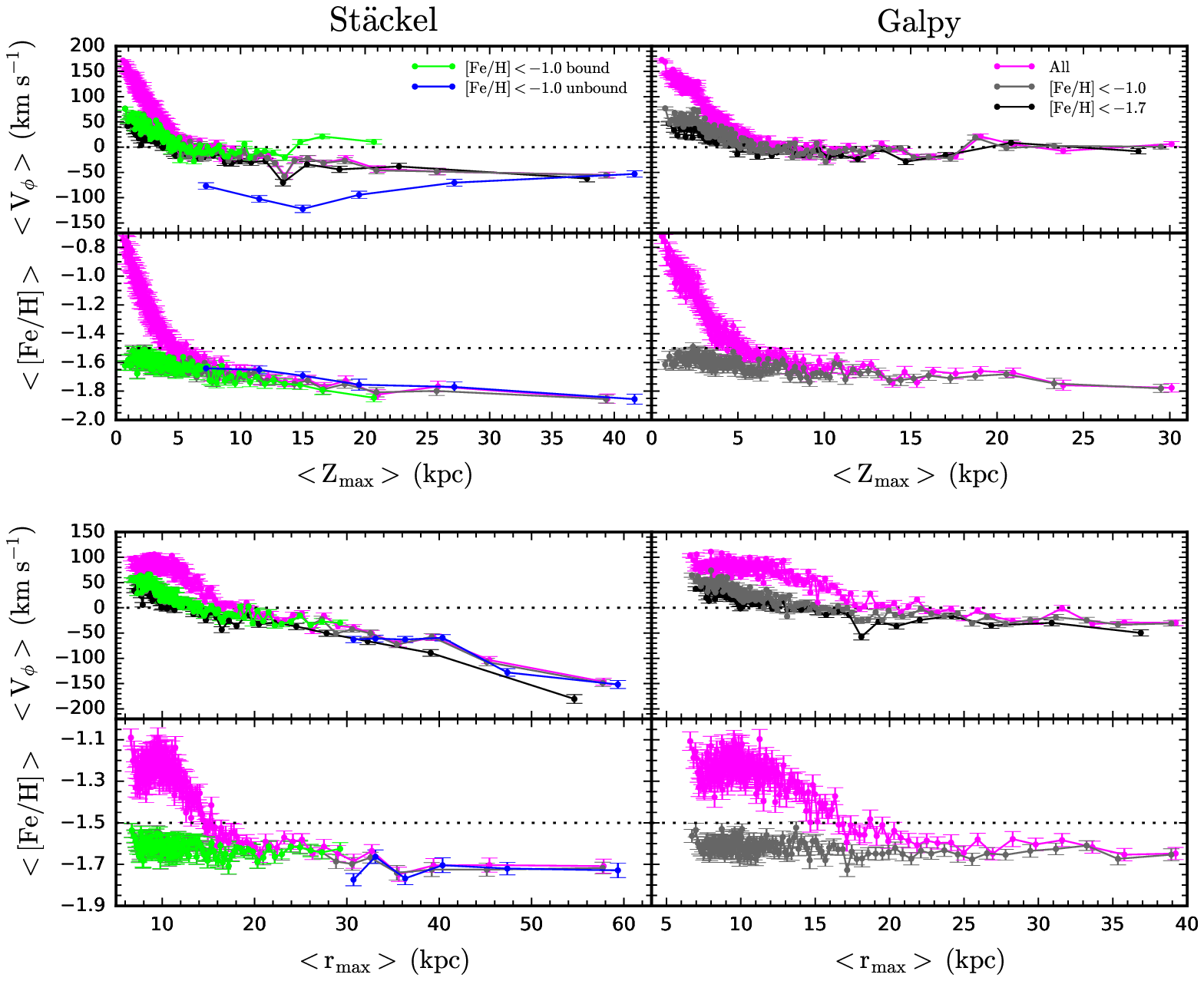}
\caption{Left panels: profiles of mean rotational velocity
($\langle$\vphi$\rangle$) and mean [Fe/H] ($\langle$[Fe/H]$\rangle$), as
a function of mean \zmax\ (top two panels) and mean \rmax\ (bottom two
panels) for the \stackel\ potential. Right panels: same as in the left
panels, but for the Galpy potential. Each mean value is obtained by
passing a box of 130 stars in \zmax\ and \rmax. As denoted in the
legends, the magenta curve is for all stars, the gray curve for stars
with [Fe/H] $<$ --1.0, and the black curve for stars with
[Fe/H] $<$ --1.7. Green and blue lines in the left panels
represent trends of the mean values calculated for bound and unbound
stars from the \stackel\ potential, respectively. The error 
bar in \vphi\ is the standard error of 20 Monte Carlo 
samples (see the text in Section \ref{sec:error} for more detailed information) 
and the one in [Fe/H] is derived from 100 bootstrap resamples of 130 
stars in each bin.}
\label{figure5}
\end{figure*}

% FIGURE 6
\begin{figure*}[t]
%\figurenum{4}
\epsscale{1.0}
\plotone{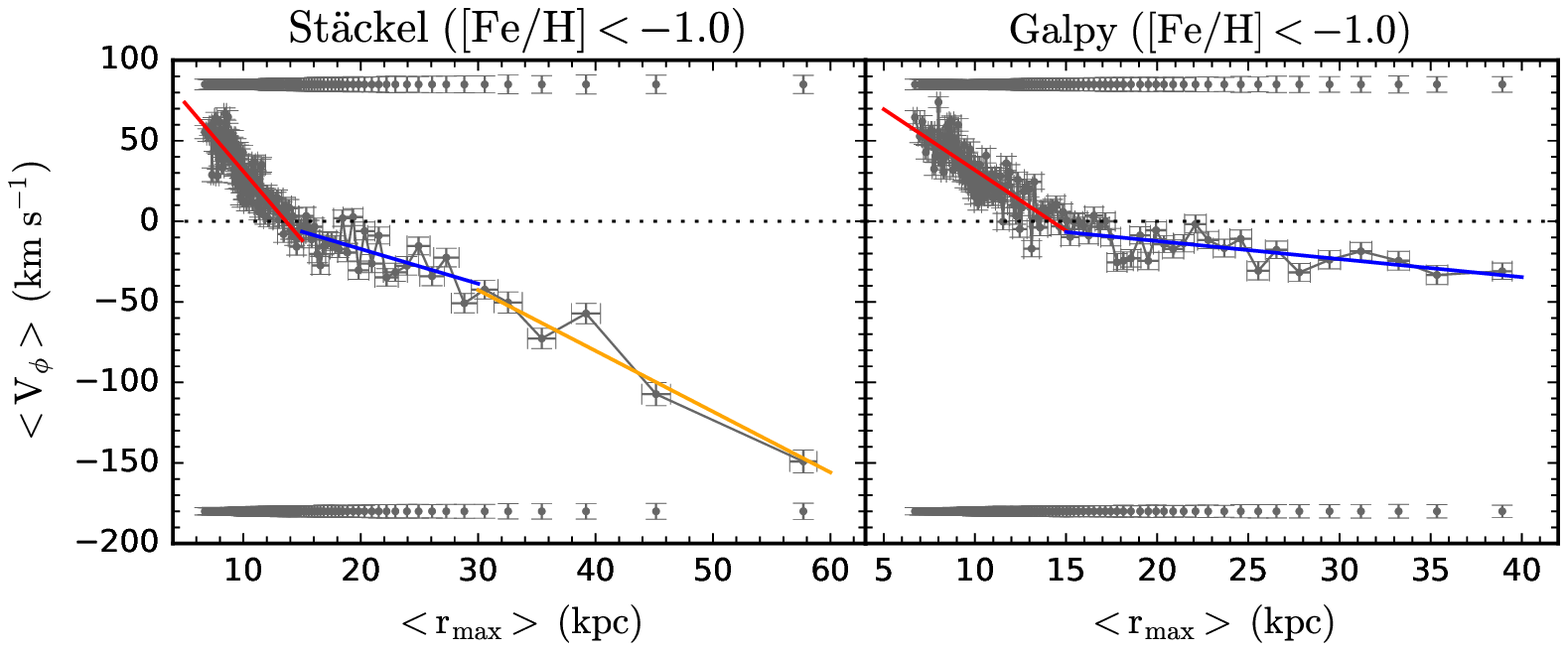}
\caption{Linear regressions on \vphi\ for stars with [Fe/H] $<$ --1.0 
in the ranges of \rmax\ $<$ 15 kpc (red line), 15 $\leq$ \rmax\ $<$ 30 kpc 
(blue line), and \rmax\ $\geq$ 30 kpc (orange line) for the \stackel\ 
potential (left panel), and \rmax\ $<$ 15 kpc (red line) and 
\rmax\ $\geq$ 15 kpc (blue line) for the Galpy potential (right panel). 
The profile of the rotational velocity is the same as for [Fe/H] $<$ --1.0 
in Figure \ref{figure5}. The upper and lower error bars are for $V_r$ and \vtheta\ 
in the \rmax\ bins, respectively. The error bars are obtained from 
20 Monte Carlo samples (see the text in Section \ref{sec:error}).}
\label{figure6}
\end{figure*}

\subsection{Differences in Halo Properties from the DR12 Sample}

In the previous section, we demonstrated that the distribution of stars
changes with \zmax, depending on the assumed potential. Accordingly, so
do the \vphi\ distributions and the MDFs. As a result, the transition
region between the inner and outer halos inferred from the shift of the
distribution of \vphi\ and [Fe/H] over \zmax\ occurs in different 
location for these potentials. Here, we examine these characteristics in
more detail with a larger extended sample from the SDSS DR12. 

Figure \ref{figure4} presents a logarithmic density map of \vt\ versus \zmax\ (top
panels) and \rmax\ (bottom panels) for the \stackel\ (left panels) and
Galpy (right panels) potentials. In this figure, the red dots are stars
in the range \zmax\ $>$ 15 kpc in the Galpy potential and in the region \zmax\ $<$
15 kpc in the \stackel\ potential. From the top panels of the figure, we note that
the displacement of stars in the Galpy potential generally reaches
larger \zmax\ ($\sim$ 35 kpc) than for the \stackel\ potential. One notable feature is that the
stars around \vt\ $\sim$ 100 \kms\ in the Galpy potential exhibit remarkable
excursions to higher \zmax\ compared to the \stackel\ potential. These
stars are mostly found close to $\alpha$ = $90^{\circ}$, as
can be inferred from the right panels of Figure \ref{figure1}, and have
larger ${V_{\rm \theta}}$ velocity components. Thus, stars
with larger ${V_{\rm \theta}}$ reach higher \zmax. Here, as $\alpha \sim
90^{\circ}$ and \vt\ $\sim$ 100 \kms, we can infer from the equation given in
Section~\ref{sec:orbit} that \vtheta\ $\sim$ 100 \kms\ and \vphi\ $\sim$ 0 \kms. 
Hence, the distribution of stars between both
potentials with \zmax\ changes primarily due to those stars.
For example, the stars shown as red dots in Figure \ref{figure4} in the region of 
\zmax\ $<$ 15 kpc in the \stackel\ potential reach up to \zmax\ = 30 kpc and many of them 
are located in the range 15 $<$ \zmax\ $<$ 25 kpc in the Galpy potential. 
As seen in the middle panels of Figure \ref{figure3},  
stars with 5 $\leq$ \zmax\ $<$ 15 kpc in the \stackel\ potential are dominated by objects 
with [Fe/H] $>$ --1.7. Because these stars extend up to \zmax\ = 20 kpc in the 
Galpy potential, the shift of the MDF to lower [Fe/H] $<$ --1.7 occurs above \zmax\ = 20 kpc 
in the Galpy potential. By way of contrast, there is no significant variation in the overall 
distribution of stars over \rmax\ between the two potentials, as can be seen in the 
bottom panels of Figure \ref{figure4}. 

Figure \ref{figure5} shows profiles of the mean rotational velocity
($\langle$\vphi$\rangle$) and mean [Fe/H] ($\langle$[Fe/H]$\rangle$), as
a function of mean \zmax\ (top two panels) and mean \rmax\ (bottom two
panels) for the \stackel\ potential in the left panels, and for the
Galpy potential in the right panels. Average values are calculated by
passing a box of 130 stars in \zmax\ and \rmax. The error 
bar in \vphi\ is obtained from 20 Monte Carlo samples (see the text in 
Section \ref{sec:error} for more detailed information); the one in [Fe/H] 
is the standard deviation of 100 bootstrap resamples of 130 stars in each bin. 
The magenta curve is for all stars, the gray curve for stars
with [Fe/H] $<$ --1.0, and the black curve from stars with [Fe/H] $<$ --1.7.
The green and blue lines in the left panels denote
trends of the mean values obtained from bound and unbound stars in the
Galpy potential, respectively.

From inspection of the magenta profiles, as a function of \zmax, for the
\stackel\ potential, we can infer that a transition in both \vphi\ and \feh\ occurs
around \zmax\ = 15 kpc. At the largest \zmax\ distance,
$\langle$\vphi$\rangle$ = --60 \kms\ and $\langle$[Fe/H]$\rangle$ =
--1.9. On the other hand, for the Galpy potential, the transition
appears to exist only for [Fe/H] above \zmax\ = 20 kpc, where
$\langle$[Fe/H]$\rangle$ = --1.8; the mean rotation velocity exhibits a
flat (nearly zero) behavior. 

In Section \ref{sec:DR7}, it was shown that the retrograde motions of stars in the
\stackel\ potential are almost independent of \zmax. This behavior 
is investigated in detail by looking into the left column of panels 
in Figure \ref{figure5}, which show the mean rotational velocity and
[Fe/H] for bound stars with a green line and for unbound stars with a
blue line. Inspection of these panels reveals that the unbound stars
show a retrograde motion at all \zmax, whereas the bound stars exhibit a
prograde motion below \zmax\ = 5 kpc and almost zero net rotation
velocity beyond that distance. We also note that the mean [Fe/H]
profiles differ between the two groups as well. Because this distinction
arises due to differences in the energy of the stellar orbits between
the bound and unbound stars, we suggest that \rmax, which is more
correlated with energy, is the more appropriate parameter to separate
the IHP from the OHP, identifying stellar populations with distinct
chemistry and kinematics in the Galactic halo. 
 
The lower left-hand panels of Figure \ref{figure5} show the mean
rotation velocity and [Fe/H], as a function of \rmax, from the \stackel\
potential; there is a clearly different behavior seen between the bound
(green line) and unbound (blue line) stars with [Fe/H] $<$ --1.0 at
\rmax\ = 30 kpc. The mean rotation velocity of bound stars shows a
prograde motion in the region \rmax\ $<$ 15 kpc, becomes less than zero
at \rmax\ = 15~kpc, and decreases slightly (to \vphi\ $\sim$ --30 \kms)
up to \rmax\ = 30 kpc. The mean rotational velocity of unbound stars
declines noticeably with increasing \rmax\ from \rmax\ = 30 kpc down to
\vphi\ $\sim$ --150 \kms\ at \rmax\ = 60 kpc. The mean [Fe/H] profile also 
exhibits an abrupt change around \rmax\ = 30 kpc, exhibiting [Fe/H] =
--1.7 above this distance.

In contrast, the mean rotation velocity (gray symbols in the right
panels of Figure \ref{figure5}) of the stars with [Fe/H] $<$ --1.0 in
the Galpy potential decreases only a little beyond \rmax\ = 15 kpc, and
remains almost flat at \vphi\ $\sim$ --30 \kms, which is similar to the
\stackel\ potential (green dots in the left panel) up to that distance.
Additionally, it is interesting to see that there is no change in the mean
[Fe/H] with increasing \rmax.

To further clarify the transition of \vphi\ with \rmax, we divided 
\rmax\ into the ranges of \rmax\ $<$ 15 kpc, 15 $\leq$ \rmax\ $<$ 30 kpc,
and \rmax\ $\geq$ 30 kpc for the \stackel\ potential, and \rmax\ $<$ 15 kpc, 
\rmax\ $\geq$ 15 kpc for the Galpy potential, based on inspection of 
Figure \ref{figure5}. We then applied linear regressions of \vphi\ 
for stars with [Fe/H] $<$ --1.0 in each region, as shown in Figure \ref{figure6}.
The profile of the rotational velocity is the same as for [Fe/H] $<$ --1.0 
in Figure \ref{figure5}. We obtained the following means and gradients
for the \stackel\ potential: 
$\langle$\vphi$\rangle = 29.9\pm 0.4~\rm km~s^{-1}$, $\Delta V_{\rm \phi}/\Delta r_{\rm max}$ = 
$-8.5\pm 0.4$ km s$^{-1}$ kpc $^{-1}$ for \rmax\ $<$ 15 kpc; 
$\langle$\vphi$\rangle = -16.8\pm 1.1~\rm km~s^{-1}$, 
$\Delta V_{\rm \phi}/\Delta r_{\rm max}$ = $-2.2\pm 0.5$ km s$^{-1}$ kpc $^{-1}$ 
for 15 $\leq$ \rmax\ $<$ 30 kpc; and $ \langle$\vphi$\rangle = -83.8\pm 2.8~\rm km~s^{-1}$, 
and $\Delta V_{\rm \phi}/\Delta r_{\rm max}$ = $-3.8\pm 0.5$ km s$^{-1}$ kpc $^{-1}$ 
for \rmax\ $\geq$ 30 kpc. For the Galpy potential, we obtained: 
$\langle$\vphi$\rangle = 31.3\pm 0.4~\rm km~s^{-1}$, 
$\Delta V_{\rm \phi}/\Delta r_{\rm max}$ = $-7.5\pm 0.4$ km s$^{-1}$ kpc $^{-1}$ 
for \rmax\ $<$ 15 kpc; and $ \langle$\vphi$\rangle = -14.6\pm 1.0~\rm km~s^{-1}$, 
$\Delta V_{\rm \phi}/\Delta r_{\rm max}$ = $-1.1\pm 0.3$ km s$^{-1}$ kpc $^{-1}$ 
for \rmax\ $\geq$ 15 kpc. Figure \ref{figure6} shows a clear discontinuity in the 
trend of \vphi\ at \rmax\ $\geq$ 30 kpc for the \stackel\ potential, but
we do not see such a trend for the Galpy potential. In Figure
\ref{figure6}, the upper and lower error bars for $V_r$ and \vtheta\ in
\rmax\ bins are shown for reference. The error bars of \rmax, \vphi,
$V_r$, and \vtheta\ are the standard deviations for 20 Monte Carlo samples
in each bin, respectively (see the text in Section \ref{sec:error}).

% FIGURE 7
\begin{figure*} %[!t]
\epsscale{1.15}
\plotone{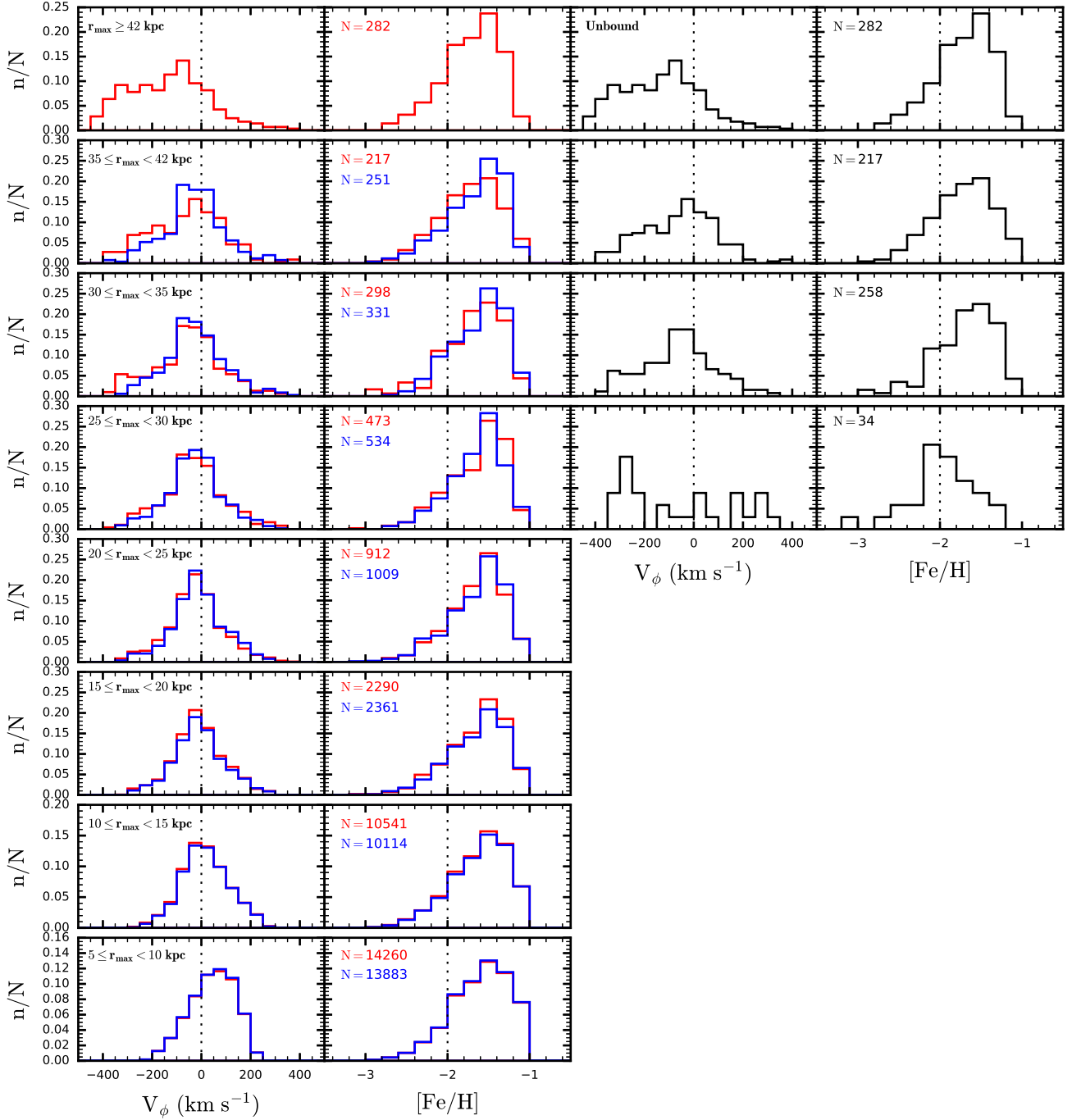}
\caption{Distribution of rotational velocities and metallicities for all
stars (first and second columns of panels) and for unbound 
stars in the Galpy potential (third and fourth columns). Here, $N$ is the 
total number of stars in each panel. Only stars with [Fe/H] $<$ --1.0 are 
considered. Histograms are shown in red and blue lines for the \stackel\ 
and the Galpy potentials, respectively.}
\label{figure7}
\end{figure*}

% FIGURE 8
\begin{figure*} %[!t]
%\figurenum{6}
\epsscale{1.15}
\plotone{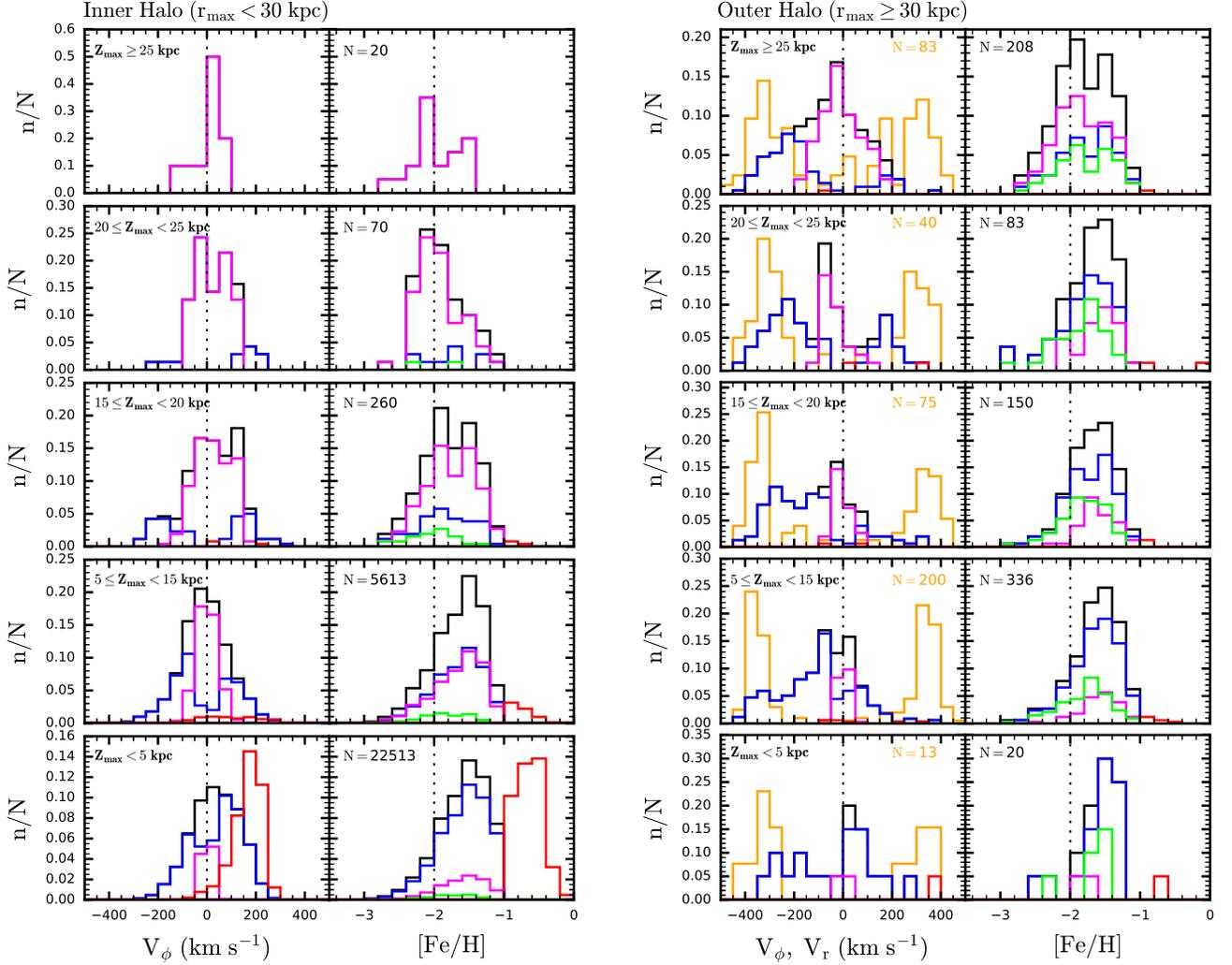}
\caption{Distribution of rotation velocities and MDFs for the IHP 
(left two columns) and the OHP (right two columns), separated at \rmax\ =
30 kpc from the \stackel\ potential, in various \zmax\ bins, as listed in
each panel. Here, $N$ is the total number of stars in each bin. Only stars
with [Fe/H] $<$ --1.0 are taken into account. The magenta histogram is
for stars with a large orbital inclination ($i > 60^{\circ}$), the blue
for the stars with small inclinations ($i < 60^{\circ}$), and black is
for all stars. The stars in the range \vphi\ $<$ --150 \kms\ are
represented by green lines in the MDFs. The red histogram indicates the
likely thick-disk stars ([Fe/H] $\geq$ --1.0) in our sample. The orange 
histogram in the first column of the right-two column panels 
is the $V_{\rm r}$ distribution of the relatively more metal-rich ([Fe/H] $>$ --1.7) stars 
in the OHP.}
\label{figure8}
\end{figure*}

% FIGURE 9
\begin{figure*} %[!t]
\epsscale{1.15}
\plotone{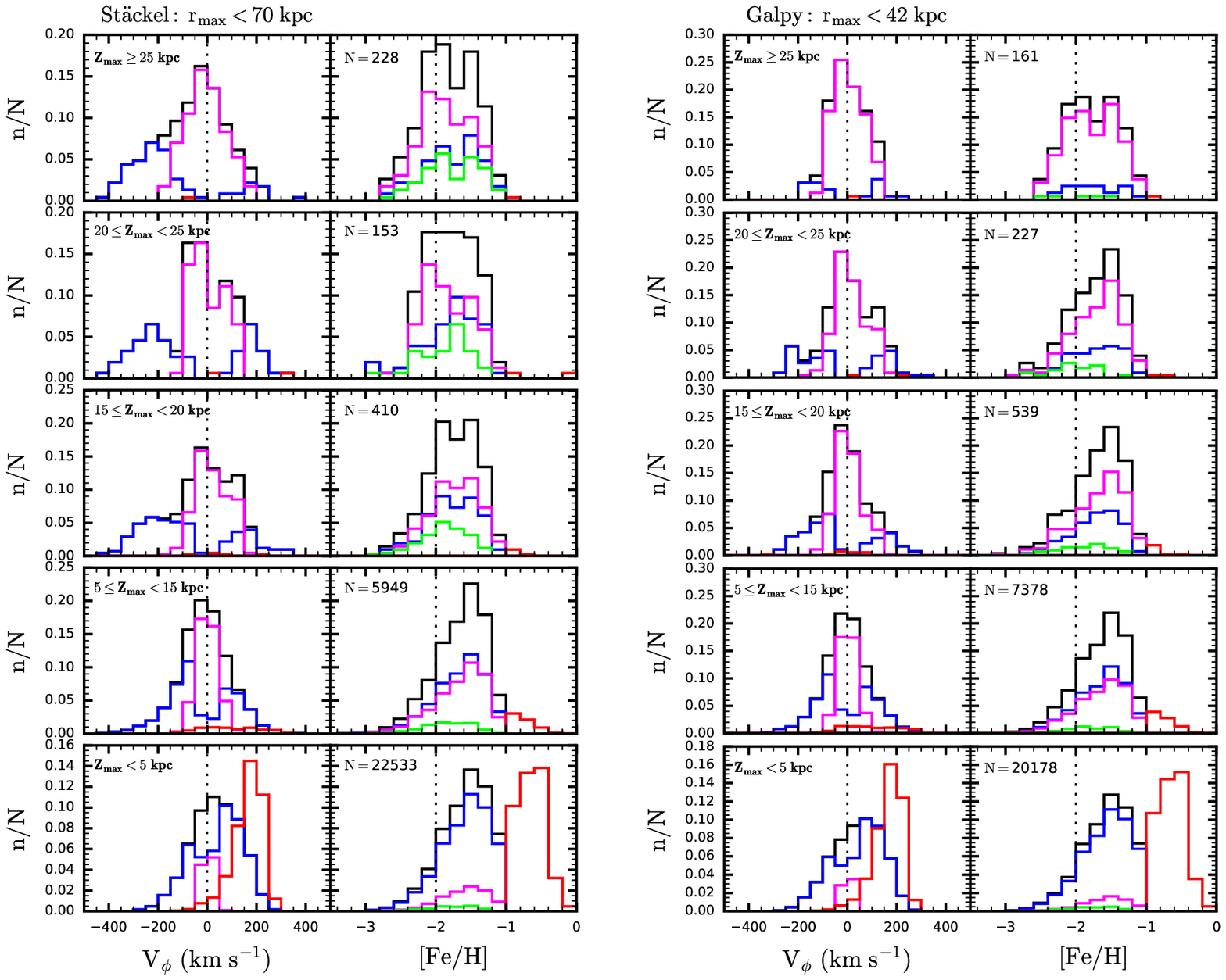}
\caption{Same as in Figure \ref{figure8}, but for the \stackel\ (left two columns) and 
the Galpy potential (right two columns), without separating the
stars into the IHP and OHP components.}
\label{figure9}
\end{figure*}

In order to confirm the results from consideration of the transition
region described above, Figure \ref{figure7} presents the distributions
of rotation velocities and metallicities for stars with [Fe/H] $<$ --1.0
in different \rmax\ bins. In the first column of panels, the peak in the
distribution of rotation velocities from the \stackel\ potential (red
line) is prograde in the range of \rmax\ $<$ 15 kpc, with a small shift
to retrograde motion from \rmax\ = 15 to 30 kpc. However, the fraction
of stars with large retrograde motions (\vphi\ $<$ --200 \kms) increases
above \rmax\ = 30 kpc, and the majority of stars in the region of \rmax\
$\geq$ 42 kpc possess retrograde motions. This trend is due primarily to
the unbound stars in the Galpy potential, as can be seen in the third
column of panels in Figure \ref{figure7}. In accordance with this
behavior, we note in the second column of panels that the fraction of
metal-poor stars becomes relatively larger in the range of \rmax\ $>$ 30
kpc, even though the peak of the distribution is still at [Fe/H] =
--1.5. 

By contrast, both the distributions of rotation velocities and
metallicities from the Galpy potential shown in the blue line in Figure
\ref{figure7} do not present any remarkable shifts in the first and
second column of panels. This leads to a conclusion that the halo stars
analyzed with the orbital parameters derived from the \stackel\
potential exhibit at least two components, while only one component is
evident when the Galpy potential is employed. Consequently, the
interpretation of the halo structure can be changed by the nature of the
adopted Galactic potential. Based on these results, we suggest a
transition distance of \rmax\ = 30 kpc from the inner to the outer halo
and adopt it as a reference point to distinguish the IHP from the OHP.

\subsection{Properties of the IHP and OHP}

For local stars at a given energy, the stars with a large orbital inclination and high
tangential velocity have orbital motions with high \zmax, while those
with a small orbital inclination and high tangential velocity show
rotational motions closer to the disk with low \zmax. On the other hand,
most stars with high radial velocities possess low
\zmax, regardless of their orbital inclination. As seen in Figure \ref{figure5}, the 
metallicity profile exhibits a more obvious transition with \zmax\ than
\rmax. This implies that a parent satellite galaxy (or possibly a
globular cluster) contributing metal-poor stars with high \zmax\ is
different from one donating stars with low \zmax. Thus, investigation of
the relationship between the orbital parameters (\zmax) and metallicity
in different ranges of energy or \rmax\ can provide valuable information
on the dynamical structure of the Galactic halo.

Figure \ref{figure8} shows the distribution of rotation velocities and 
metallicities of the IHP (left-two columns) and the OHP (right-two columns) 
divided at \rmax\ = 30 kpc, in different bins of \zmax, calculated from the
\stackel\ potential for the DR12 sample. The cut of \rmax\ = 30 kpc was derived based on the
change in the shape of the \vphi\ distribution for the stars in the range [Fe/H]
$<$ --1.0, as described in the previous subsection. In the figure, the
magenta histogram is for the stars with large orbital inclination ($i
> 60^{\circ}$), the blue histogram for the stars with small inclination ($i <
60^{\circ}$), and the black histogram for all stars. The green distribution in the
panels for the MDF indicates the stars in the range \vphi\ $<$ --150
\kms. The red histogram is constructed from stars with [Fe/H] $\geq$ --1.0 
in our sample. The orange histogram is the $V_{\rm r}$ 
distribution of metal-rich ([Fe/H] $>$ --1.7) stars in the OHP.

Inspection of the MDFs of the IHP in Figure \ref{figure8} reveals that 
metal-poor ([Fe/H] $<$ --1.7) stars start to dominate in the region 
above \zmax\ = 15 kpc. These metal-poor stars have relatively large orbital 
inclinations (magenta histogram) and a large dispersion in \vphi. In this 
region (\zmax\ $>$ 15 kpc), the net rotational
velocity is nearly zero, as can be seen in the left column of the panels. On
the other hand, metal-rich ([Fe/H] $>$ --1.7) stars dominate below \zmax\ = 15 kpc. 
The number of stars with high orbital inclinations (magenta histogram) is
similar to the number of stars with small orbital inclinations (blue
histogram) in the range 5 $\leq$ \zmax\ $<$ 15 kpc, while most stars in
the region of \zmax\ $<$ 5 kpc have disk-like motions, as can be seen
from the red histogram constructed from stars with [Fe/H] $\geq$ --1.0;
these are likely members of the metal-weak thick-disk population. The
distribution of the rotation velocity in the range of 5 $\leq$ \zmax\
$<$ 15 kpc is symmetric about \vphi\ = 0, while the number of stars with
prograde motions increases in the region of \zmax\ $<$ 5 kpc, compared
to the numbers of stars with retrograde motions. 

% FIGURE 10
\begin{figure*} [!t]
\epsscale{1.15}
\plotone{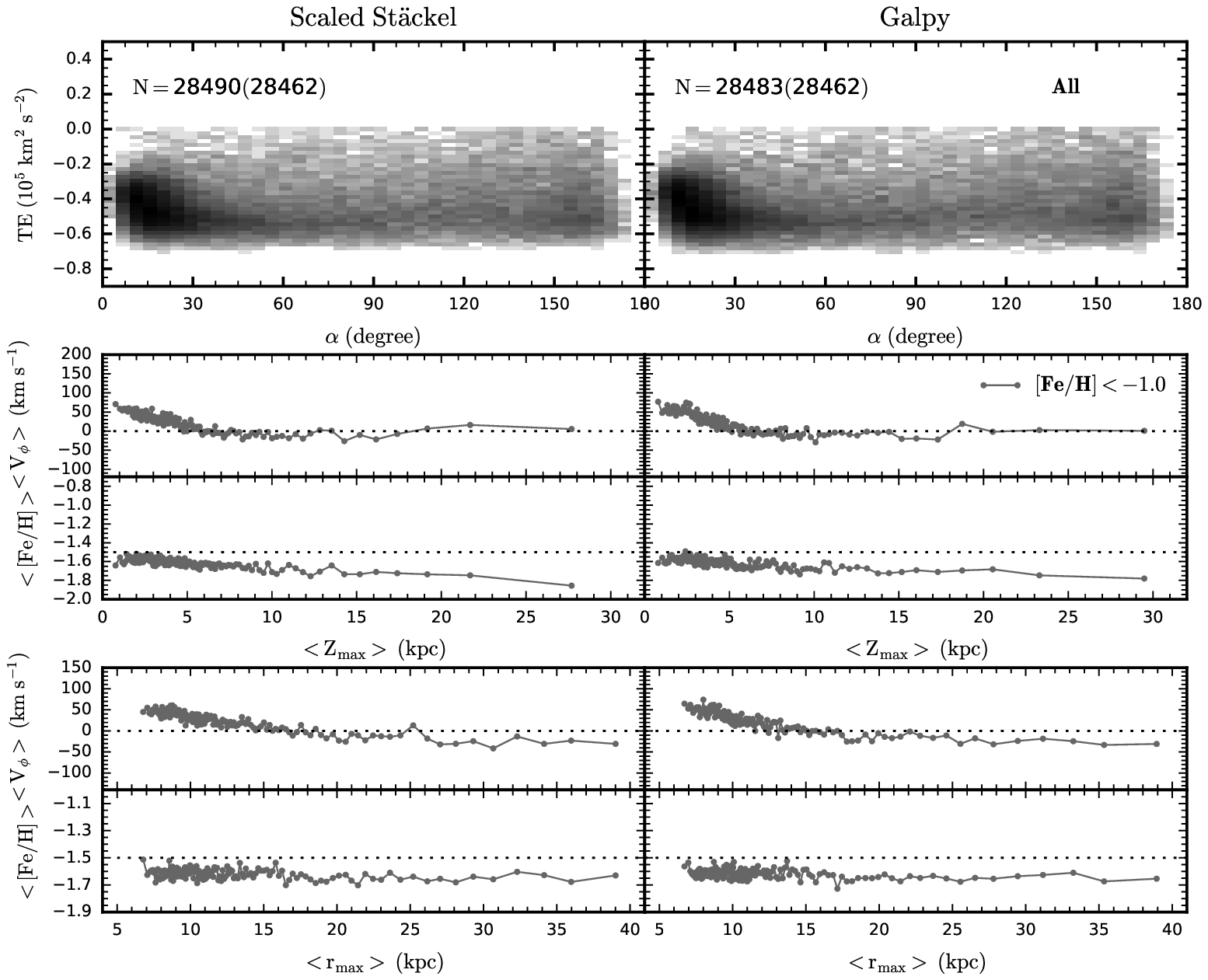}
\caption{Left panels: distribution of total energy (TE) versus ${\alpha}$ for all 
stars (top panel) and profiles of mean rotational velocity and mean [Fe/H] for stars 
with [Fe/H] $<$ --1.0, as a function of mean \zmax\ (middle two panels) and mean 
\rmax\ (bottom two panels) for the scaled \stackel\ potential. Right panels: 
same as in the left panels, but for the Galpy potential. Numbers in the top panels indicate 
how many bound stars are in the respective potentials. Numbers in brackets indicate how many  
stars are bound in both potentials. Maps in the top panels display the 
number density (low to high from bright to dark) on a log-10 based scale for 
all stars. Each mean value is obtained by passing a box of 130 stars in \zmax\ and \rmax.}
\label{figure10}
\end{figure*}

In the OHP shown in the right two panels 
of Figure \ref{figure8}, it is clear that most stars exhibit retrograde
rotation, with \vphi\ $<$ --150 \kms, independent of \zmax, although
there are some stars with prograde motions. A large portion of these
stars have low orbital inclinations (blue histogram). In the range of
\zmax\ $\geq$ 25 kpc, the stars near \vphi\ = 0 have high orbital
inclinations, while in the range of \zmax\ $<$ 25 kpc, the stars exhibit
a large \vphi\ velocity dispersion regardless of their \zmax\ cuts.

It is also interesting to note that, in the MDFs of the OHP, there is a
greater fraction of metal-poor ([Fe/H] $<$ --1.7) stars with large
retrograde motions (\vphi\ $<$ --150 \kms) than metal-rich ([Fe/H] $>$
--1.7) stars, regardless of \zmax\ (see the green histograms of the
right two columns of Figure \ref{figure8}). We also notice that
metal-poor stars become more dominant in the region of \zmax\ $\geq$ 25
kpc, while metal-rich stars become more populous in the range of \zmax\
$<$ 25 kpc. This feature is slightly different from the IHP, which shows
that the dominance of the metal-poor stars appear to occur at \zmax\ =
15 kpc. 

The different spatial distributions of the 
metal-poor stars between the IHP and OHP can be understood as follows.
As the most metal-poor stars in the OHP have primarily high inclination
with high tangential orbits and high retrograde orbits, they can easily
reach above \zmax\ = 25 kpc, resulting in their occupation of the region
\zmax\ $\geq$ 25 kpc. On the other hand, because the more metal-rich stars
have strong radial (see the orange histograms in the first of the
right-two columns of Figure \ref{figure8}) and retrograde motions, they
do not reach high \zmax---rather, they reside at lower \zmax\ ($<$ 25 kpc).
Generally, at a given energy, stars with high $V_{\theta}$ reach
higher \zmax\ than those with high $V_{\rm r}$; reaching a distance
of \zmax\ $>$ 25 kpc requires more energy. Hence, the OHP stars with
high energy and the IHP stars with low energy represent more metal-poor
stars beyond \zmax\ = 25 and 15 kpc, respectively. 
  
Figure \ref{figure9} shows the distribution of rotation velocities
and metallicities for the \stackel\ (left two columns) and the Galpy
potential (right two columns), {\it without} dividing stars into the IHP
and OHP, for the purpose of comparison to Figure \ref{figure8}. As
shown in Figure \ref{figure9}, the \stackel\ potential exhibits a
retrograde motion, with metal-poor stars dominating beyond 
\zmax\ = 15 kpc, whereas the Galpy potential only indicates the shift in the 
MDF above \zmax\ $>$ 25 kpc. We re-emphasize that the reason for the
lack of retrograde motions in the Galpy potential, compared to the
\stackel\ potential, is not because the stars near $\alpha = 90^{\circ}$
move up to higher \zmax, but rather because the stars with large retrograde orbits are
not bound to the potential.

% FIGURE 11
\begin{figure*} [!t]
\epsscale{1.0}
\plotone{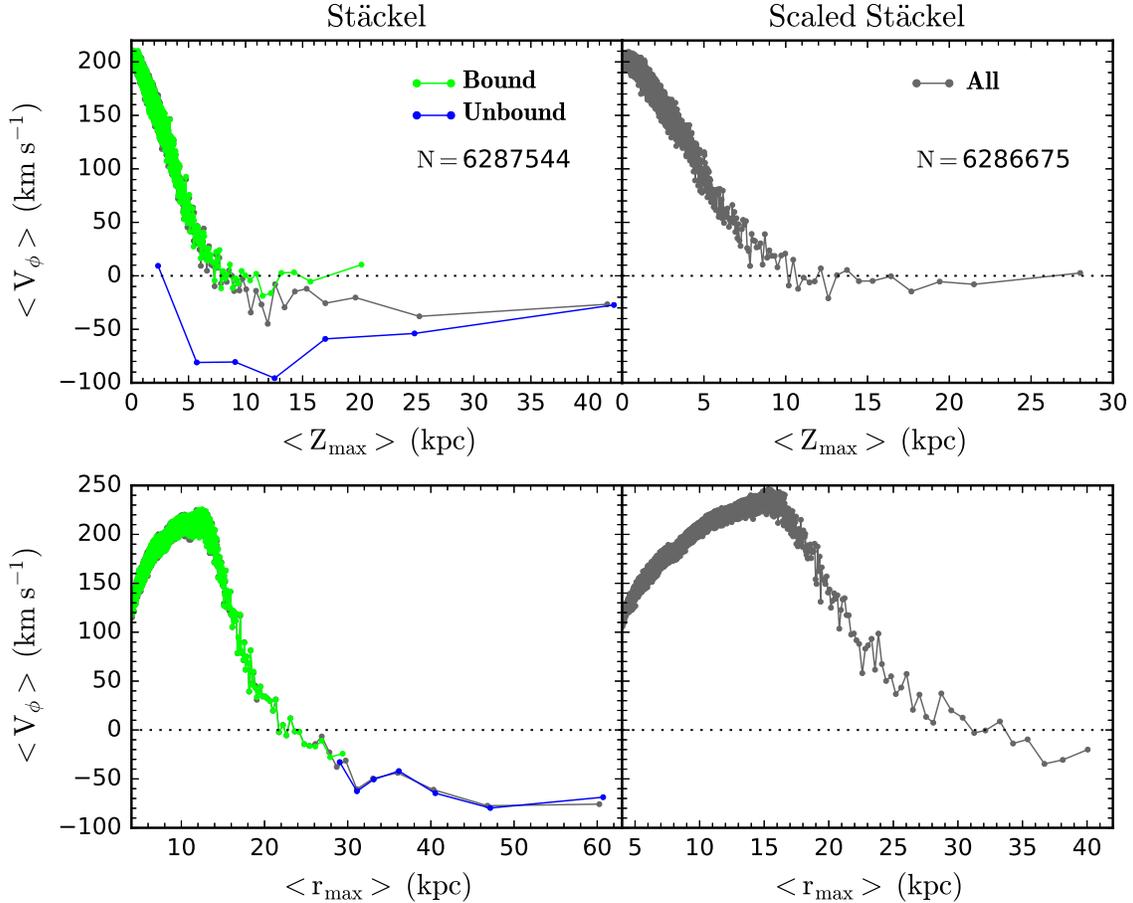}
\caption{Profiles of mean rotational velocity ($\langle$\vphi$\rangle$) for stars 
in the RVS sample drawn from $Gaia$ DR2.
Left panels: profiles of mean \vphi, as a function of mean \zmax\ (top panel)
and mean \rmax\ (bottom panel), for the \stackel\ potential. 
Right panels: same as in the left panels, but for the scaled \stackel\ potential. 
Each mean value is obtained by passing a box of 130 stars in \zmax\ and \rmax.
Here, $N$ is the total number of bound stars in each potential. As denoted in the legends, 
the gray curve is for all stars. The green and blue lines in the left panels represent 
trends of the mean values calculated for bound and unbound stars from the scaled \stackel\ 
potential, respectively.}
\label{figure11}
\end{figure*}

% FIGURE 12
\begin{figure*}[t]
%\centering
%\figurenum{1}
\epsscale{1.0}
\plotone{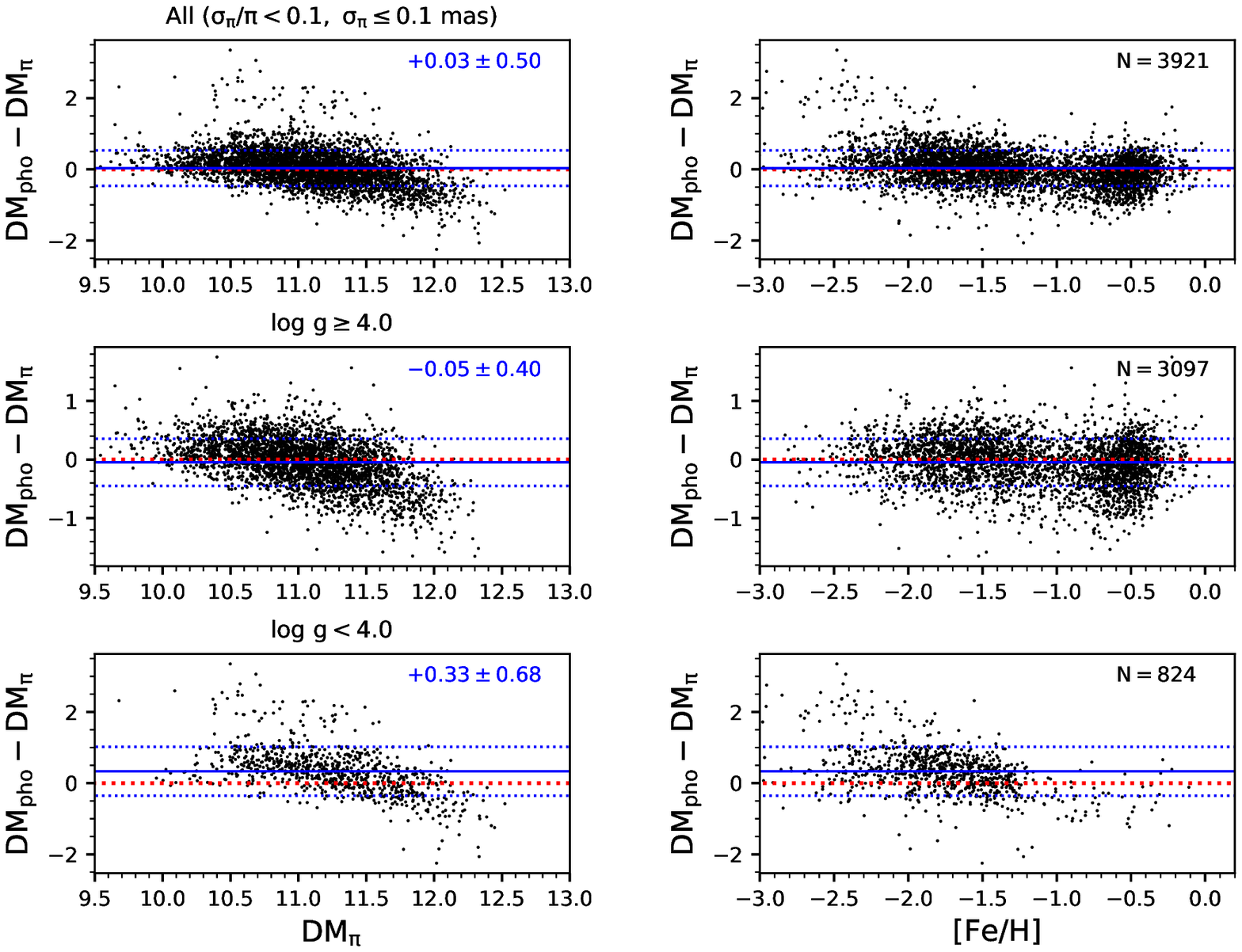}
\caption{Comparison of our photometric distances (DM$_{\rm pho}$) and 
those (DM$_{\rm \pi}$) of the $Gaia$ DR2. Top panels are for all stars, middle for 
stars with $\log g \geq 4.0$, and bottom for those with $\log g < 4.0$ as a function
of distance modulus (left) and metallicity (right). We used distances corrected by 
the zero-point offset of --0.029 mas and considered only stars in the ranges
of $\sigma_\pi / \pi < 0.1$ and $\sigma_\pi \leq 0.1 ~\mathrm{mas}$. Numbers in the top right areas
of the left and right panels are the means and standard deviations of the differences in distance modulus 
and the number of stars in each panel.
}
\label{figure12}
\end{figure*}

\subsection{Halo Properties in the Scaled St$\ddot{a}$ckel Potential
Compared to the Galpy Potential} \label{sec:scaled potential}

We also carried out an exercise of rescaling the mass distribution of the \stackel\ 
potential to approximately equal that of the Galpy potential, in order
to check whether we obtain similar halo properties from the rescaled
\stackel\ and Galpy potentials. For this exercise, we first set the
tidal cutoff radius of the \stackel\ potential to 42 kpc, based on the
upper limit of $< 42$ kpc of \rmax\ from the Galpy potential, and the
disk mass to $M_{\rm d} = 7.3 \times 10^{10}~M_{\odot}$, which is the
sum of the bulge and disk masses from the Galpy potential. We then
determined the central density value of $\rho_0 = 1.99 \times 10^7
~M_{\odot}~\rm{kpc}^{-3}$, by adjusting that of the halo to make the
depth of the scaled \stackel\ potential approximately equal to that of
the Galpy potential. 

Figure \ref{figure10} shows the distribution of 
TE versus ${\alpha}$ for all stars (top panel), along with profiles of the
mean rotational velocity ($\langle$\vphi$\rangle$) and mean [Fe/H]
($\langle$[Fe/H]$\rangle$) for stars with [Fe/H] $<$ --1.0 as a
function of mean \zmax\ (middle two panels) and mean \rmax\ (bottom two
panels) for the scaled \stackel\ potential (left panels) and for
the Galpy potential (right panels). Average values are calculated
by passing a box of 130 stars in \zmax\ and \rmax. The figure clearly
demonstrates that we obtain nearly identical halo properties from both the
scaled \stackel\ and Galpy potentials. This confirms that the different
characteristics of the Galactic halo between the \stackel\ and Galpy
potentials stem primarily from the different potential depths.

\subsection{Retrograde Motion Signature from the Radial Velocity
Spectrometer Sample of $Gaia$ DR2} \label{sec:rvsample}

In order to investigate the reality of the claimed retrograde motion in the \stackel\ and 
scaled \stackel\ potentials, we have also made use of a sample of stars
with available radial velocity measurements from the radial velocity
spectrometer (RVS) in $Gaia$ DR2.  We first selected stars satisfying ${\pi}/{\sigma}_{\pi} > 4$ 
and proper motion errors less than 1.0 $\mathrm{mas~yr^{-1}}$ in the RVS
sample from $Gaia$ DR2, using distances derived by \citet{schonrich2019}, who estimated the Bayesian 
distance for the RVS stars in $Gaia$ DR2.

Figure \ref{figure11} shows profiles of the mean rotational velocity
($\langle$\vphi$\rangle$), as a function of mean \zmax\ (top panel) 
and mean \rmax\ (bottom panel) for the \stackel\ potential in the left panels, 
and for the scaled \stackel\ potential in the right panels. Average values were 
calculated by passing a box of 130 stars in \zmax\ and \rmax. The gray curve is 
for all stars. The green and blue lines in the left panels denote trends of the 
mean values obtained from bound and unbound stars in the scaled \stackel\ potential,
respectively. Figure \ref{figure11} verifies that RVS sample of $Gaia$ 
DR2 also shows the signature of retrograde motion for \rmax\ $\geq$ 30 kpc, consistent 
with our result for the \stackel\ potential. The left column of panels in 
Figure \ref{figure11} also reveal that the rotational velocity depends more strongly 
on \rmax\ than on \zmax.

% FIGURE 13
\begin{figure*} %[!t]
\epsscale{0.9}
\plotone{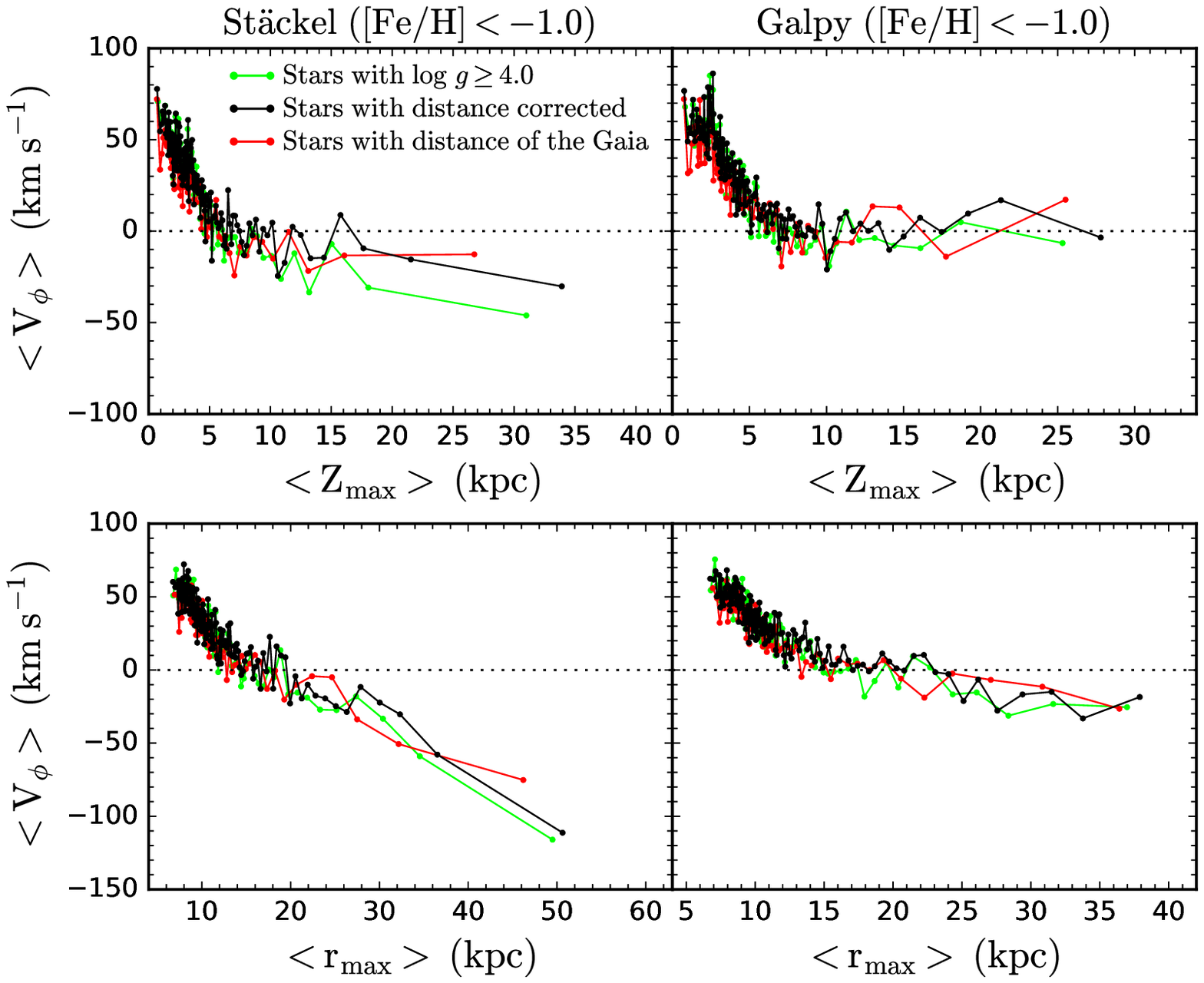}
\caption{Profiles of the mean rotational velocity ($\langle$\vphi$\rangle$), as
a function of mean \zmax\ (top two panels) and mean \rmax\ (bottom two panels) 
for three groups of stars with [Fe/H] $<$ --1.0, for the \stackel\ potential 
in the left panels, and for the Galpy potential in the right panels. 
The first sample consists of stars with $\log g \geq 4.0$, which 
have a negligible distance offset. The second sample includes stars with $\log g \geq 4.0$ and 
those with $\log g < 4.0$, corrected for the distance offset by $+$0.33 dex. 
The third group comprises stars chosen from our calibration stars, using the selection
criteria described in Section~\ref{sec:sel}, with the distance estimate
from $Gaia$ DR2, including stars that satisfy $\sigma_\pi /
\pi < 0.2$ and $\sigma_\pi \leq 0.1$ mas.
The first, second, and third groups are displayed in green, black, and red colors, respectively. Each 
mean value is obtained by passing a box of 130 stars in \zmax\ and \rmax.}
\label{figure13}
\end{figure*}

% FIGURE 14
\begin{figure*} %[!t]
\epsscale{1.15}
\plotone{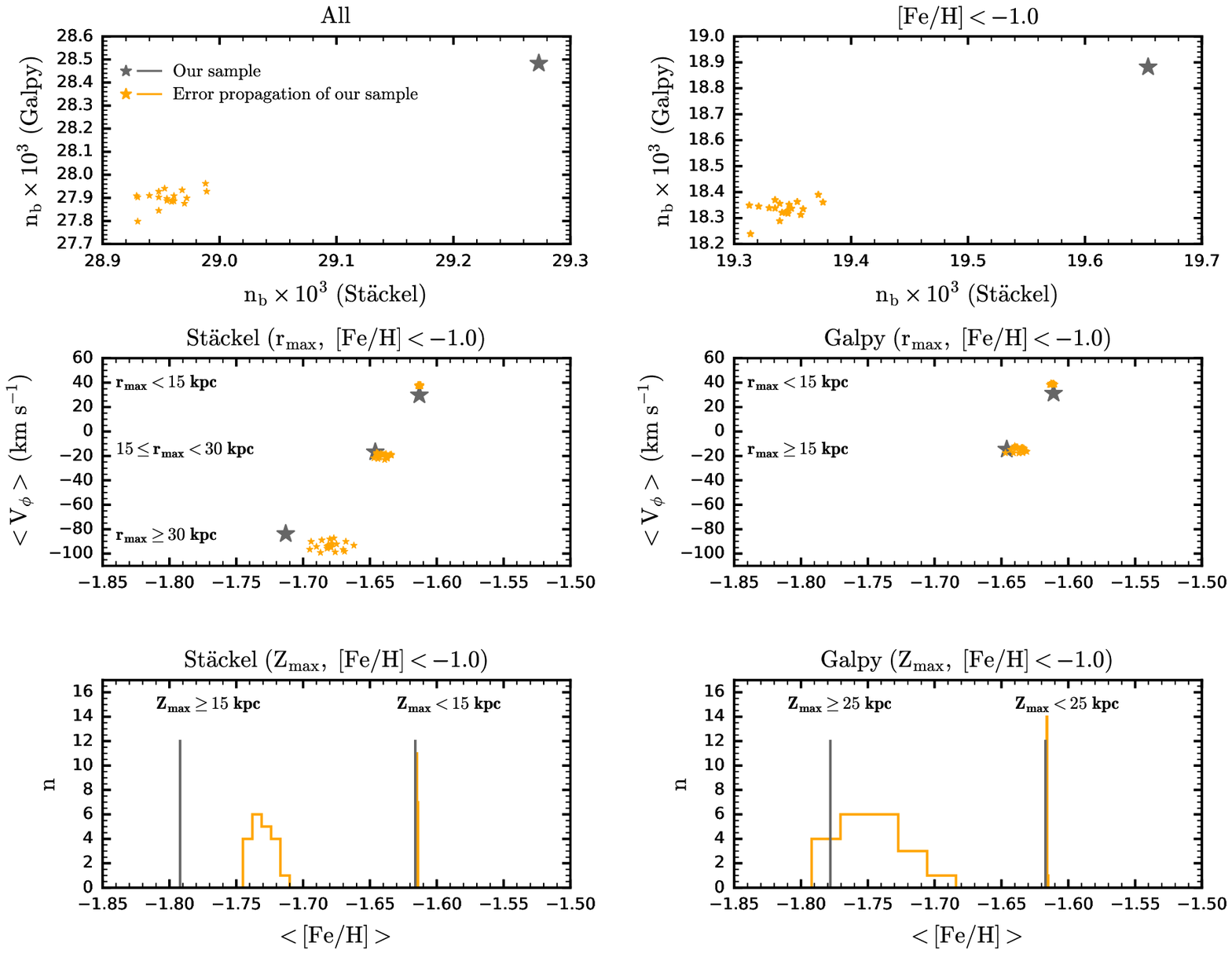}
\caption{Effects of observational uncertainties based on 20 mock samples 
obtained from a Monte Carlo simulation. Top panels: mean \vphi\ versus
mean [Fe/H] for stars with [Fe/H] $<$ --1.0 in the regions of \rmax\ $<$
15 kpc, 15 $\leq$ \rmax\ $<$ 30 kpc, and \rmax\ $\geq$ 30 kpc 
for the \stackel\ potential (left panels), and \rmax\ $<$ 15 kpc and \rmax\
$\geq$ 15 kpc for the Galpy potential (right panels). Our sample and each
of the 20 mock samples are displayed with the gray star symbols and the orange
star symbols, respectively. Bottom panels: distribution of mean [Fe/H]
for stars with [Fe/H] $<$ --1.0 in the ranges of \zmax\ $\geq$ 15 kpc and
\zmax\ $<$ 15 kpc for the \stackel\ potential (left panels), and
\zmax\ $\geq$ 25 kpc and \zmax\ $<$ 25 kpc for the Galpy
potential (right panels). Our sample and each of the 20 mock samples are displayed with the gray
solid line and the orange solid histogram, respectively.}
\label{figure14}
\end{figure*}

\section{Impacts of Distance Error and Selection Bias on Derived Orbital Parameters} \label{sec:bias}

As any systematic error on estimated measured distance and/or metallicity bias  
in target selection can affect the derived orbital parameters and any
interpretation that follows, below we have investigated these possible impacts.

\subsection{Impact of Distance Errors} \label{sec:error}

In order to check whether systematic errors exist in our distance
estimates, we matched our calibration-star sample from SDSS DR12 with stars in $Gaia$
DR2, and selected stars satisfying ${\sigma}_{\pi}/{\pi} < 0.1$
and ${\sigma}_{\pi} \leq 0.1 ~\rm{mas}$, where $\pi$ is the parallax and
${\sigma}_{\pi}$ is the uncertainty in $\pi$. Distances from $Gaia$ DR2
were derived by inverting parallaxes, corrected for the zero-point
offset of --0.029 $\rm{mas}$ (\citealt{lindegren2018}). Figure
\ref{figure12} shows the differences between our photometric distances
(DM$_{\rm pho}$) and those from $Gaia$ DR2 (DM$_{\rm \pi}$), as a function
of distance modulus (left panels) and metallicity (right panels) for: (1)
all matched stars (top panels), (2) stars with $\log g \geq 4.0$ (middle
panels), and (3) stars with $\log g < 4.0$ (bottom panels). We found a
mean offset of 0.03 dex with a scatter of 0.5 dex for all stars, --0.05
dex with a standard deviation of 0.40 dex for stars with $\log g \geq
4.0$, and an offset of 0.33 dex with a scatter of 0.68 dex for stars
with $\log g < 4.0$. As can be appreciated from inspection of the
figure, even though the mean offset becomes larger for fainter objects,
any systematic difference in the distance is very small. We also note
that there is no significant trend with metallicity.

We further examined the \vphi\ profile, as a function of 
\zmax\ and \rmax, because the different distance scale could result in different  
derived orbital parameters. First, based on Figure \ref{figure12}, we
selected three groups of stars. The first group consists of stars with
$\log g \geq 4.0$, which have a negligible distance offset. The second
contains the first group plus the stars with $\log g < 4.0$, which are
corrected for the distance offset by $+$0.33 dex. The third group
comprises stars chosen from our calibration stars, using the selection
criteria described in Section~\ref{sec:sel}, with the distance estimate
from $Gaia$ DR2, including stars that satisfy $\sigma_\pi /
\pi < 0.2$ and $\sigma_\pi \leq 0.1$ mas. The
numbers of bound stars for the \stackel\ and Galpy potential are $20,779$
and $20,400$ for the first group, $29,273$ and $28,785$ for the second
group, and $13,564$ and $13,296$ for the third group, respectively.

Figure \ref{figure13} shows profiles of the mean rotational velocity
($\langle$\vphi$\rangle$), as a function of mean \zmax\ (top two
panels) and mean \rmax\ (bottom two panels), for the three groups of stars with 
[Fe/H] $<$ --1.0 for the \stackel\ potential in the left panels, and 
for the Galpy potential in the right panels. The first, second, 
and third groups are displayed in green, black, and red colors, respectively. 
Average values are calculated by passing a box of 130 stars in \zmax\ 
and \rmax. Error bars are not plotted, for a more clear view of the trends.
The obvious retrograde motions of all groups for \rmax\ $\geq$ 30 kpc in the 
\stackel\ potential indicate that the retrograde motion of our 
sample is not due to overestimation of our distances.

To assess the impact of observational uncertainties on 
our derived orbital parameters, we also carried out a comparison of orbital parameters 
derived for our sample with those of 20 mock samples based on Monte Carlo simulations 
with an uncertainty of 20\% in the distance and quoted uncertainties 
in the radial velocity and proper motion, assuming a normal error distribution 
for our sample.

The comparison of our sample with each of the 20 mock samples is 
performed as follows. For the \stackel\ potential, we first separated \rmax\ into 
three regions, \rmax\ $<$ 15 kpc, 15 $\leq$ \rmax\ $<$ 30 kpc, and \rmax\ 
$\geq$ 30 kpc, and calculated mean \vphi\ and [Fe/H] for stars with [Fe/H] $<$ --1.0 
in each region. Similarly, we divided \zmax\ into two regions, \zmax\ $<$ 15 kpc and 
\zmax\ $\geq$ 15 kpc, and computed each mean [Fe/H] for stars with [Fe/H] $<$ --1.0. 
For the Galpy potential, we separated \rmax\ into two regions of \rmax\ $<$ 15 kpc 
and \rmax\ $\geq$ 15 kpc, and calculated the mean \vphi\ and [Fe/H] for stars with 
[Fe/H] $<$ --1.0 in each region. For \zmax, we divided into two regions, \zmax\ 
$<$ 25 kpc and \zmax\ $\geq$ 25 kpc, and derived each mean [Fe/H] for stars with 
[Fe/H] $<$ --1.0.

% FIGURE 15
\begin{figure*}[t]
%\centering
%\figurenum{1}
\epsscale{0.9}
\plotone{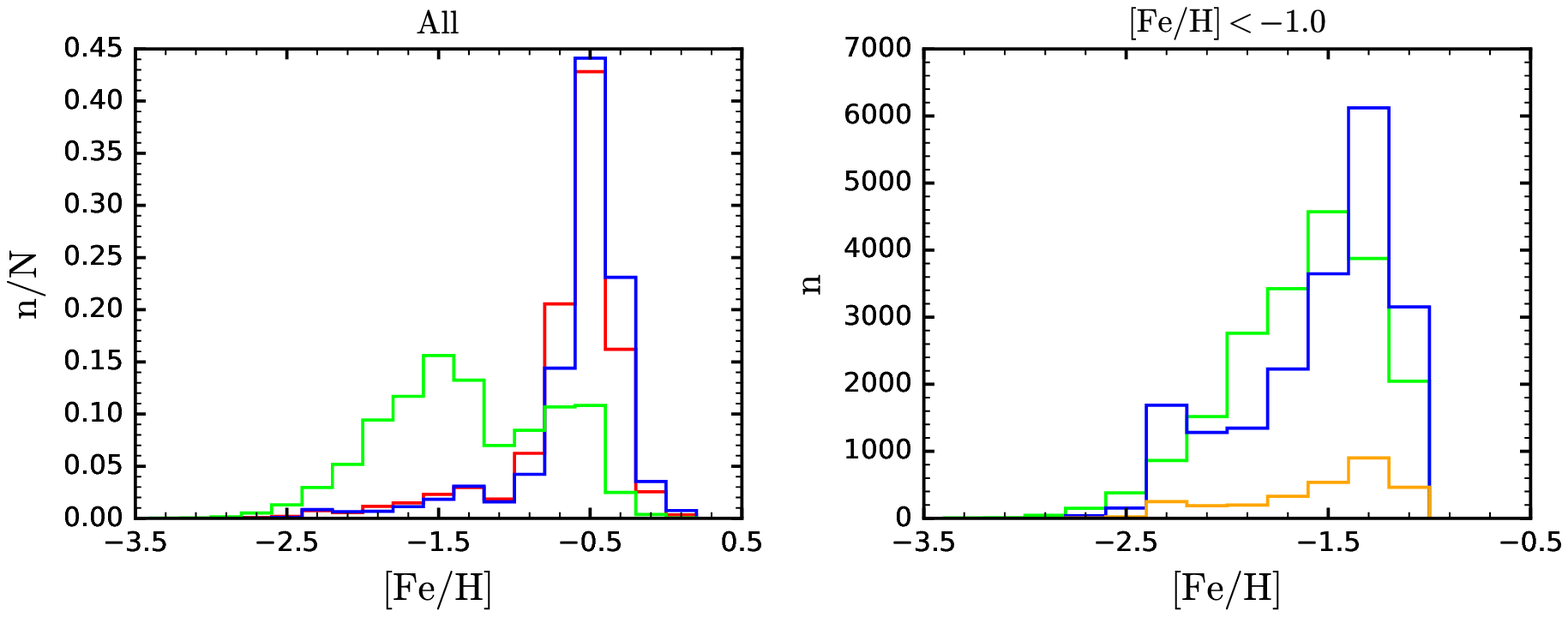}
\caption{Metallicity distribution functions of stars satisfying
our sample selection, from the SEGUE data of \citet{mints2019}.
Left panel: fractional MDFs, uncorrected (red line) and corrected
(blue line) for SEGUE target selection effects, according to
\citet{mints2019}. Right panel: the orange histogram is a scaled MDF
of the corrected MDF (the blue histogram in the left panel), which is
obtained by multiplying by the total number of stars in our sample. The
blue histogram is a rescaled MDF derived by multiplying the orange
histogram by the ratio of the number of stars in our sample ($19,654$)
to that of the scaled and corrected MDF ($2895$). Note that, in the
right panel, we only considered the stars with [Fe/H] $<$ --1.0, which
is the metallicity range of interest in this study. The green histogram
in both panels represents our sample of stars.}
\label{figure15}
\end{figure*}

Figure \ref{figure14} summarizes the comparison. 
The top panels of the figure show the numbers of bound stars in the
\stackel\ and Galpy potential, for all stars(left panel) and for only those
stars with [Fe/H] $<$ --1.0(right panel). A gray star symbol is
used for our sample, while the orange star symbols represent the 20 mock
samples. The numbers of unbound stars due to observational uncertainties
from the 20 simulated samples are, on average, $317$ and $584$ in the
\stackel\ and Galpy potential, respectively, 
for all stars, and $311$ and $548$ for stars with [Fe/H] $<$ --1.0. The
middle panels display the mean \vphi\ versus the mean [Fe/H] for stars
with [Fe/H] $<$ --1.0, in the regions \rmax\ $<$ 15 kpc, 15 $\leq$
\rmax\ $<$ 30 kpc, and \rmax\ $\geq$ 30 kpc, in the left panel for the
\stackel\ potential, and \rmax\ $<$ 15 kpc and \rmax\ $\geq$ 15 kpc, in
the right panel for the Galpy potential. The gray star symbol is for our
sample, while the orange star symbols denote the 20 mock samples. Note
that as \rmax\ increases, the difference in \vphi\ and [Fe/H] between
our original sample and the simulated samples becomes larger, but not
dramatically so.

The bottom panels represent the distribution of mean [Fe/H] of stars with 
[Fe/H] $<$ --1.0, in the ranges of \zmax\ $\geq$ 15 kpc and \zmax\ $<$
15 kpc for the \stackel\ potential (left), and \zmax\ $\geq$
25 kpc and \zmax\ $<$ 25 kpc for the Galpy
potential (right). The gray solid line indicates our sample, and the orange solid
histogram is for the 20 mock samples. In this case, similar to \rmax,
the difference between our sample and the simulated samples is larger
for larger \zmax\ for both potentials, but the difference is less than
0.1 dex. Thus, from inspection of Figure \ref{figure14}, we find that
there are no large impacts on our results due to uncertainties in the
various derived kinematic quantities.  

\subsection{Impact of Target Selection Effects} \label{sec:selection}

In order to check the impact of the target selection effect 
in our sample on the derived chemo-kinematic features reported in
Section \ref{sec:results}, we made use of the selection function for
SEGUE data provided by \citet{mints2019}. The left panel of Figure
\ref{figure15} shows the fractional MDFs, both uncorrected (red line) and
corrected (blue line) for SEGUE target selection effects. These 
MDFs are constructed with the SEGUE stars (\citealt{mints2019}) 
that satisfy our stated sampling criteria in Section~\ref{sec:sel}. We 
then assume that the blue histogram shown in the left panel of 
Figure \ref{figure15} represents the MDF that we will scale to, as 
described below. 

Derivation of the selection-corrected (unbiased) MDF for our
sample stars proceeds as follows. First, we scaled the fractional
unbiased MDF (blue histogram in the left panel of Figure
\ref{figure15}) by multiplying by the total number of stars in our
sample. This scaled MDF is shown as the orange histogram in the right
panel of Figure \ref{figure15}. We then rescaled this MDF (the orange
histogram) by multiplying by the ratio of the number of stars of our
sample ($19,654$; the green histogram) to that of the scaled unbiased
MDF ($2895$; the orange histogram). This rescaled MDF is shown as the blue
histogram in the right panel of Figure \ref{figure15}. Note that we have
restricted the sample to only include the stars with [Fe/H] $<$ --1.0,
which is the metallicity regime of interest for this study.
Finally, for each metallicity bin, we calculated the difference in the
number of stars between the MDF of our uncorrected sample (green
histogram in the right panel of Figure \ref{figure15}) of our sample and
the unbiased MDF (blue histogram in the right panel of Figure
\ref{figure15}), and randomly draw stars from our sample as many times as required to
add to (or subtract from) our sample in order to match the differences
in the number of stars between the our uncorrected MDF and the unbiased
MDF in each metallicity bin. We carried out this procedure 100 times to
obtain 100 different samples.

% FIGURE 16
\begin{figure*} %[!t]
\epsscale{1.15}
\plotone{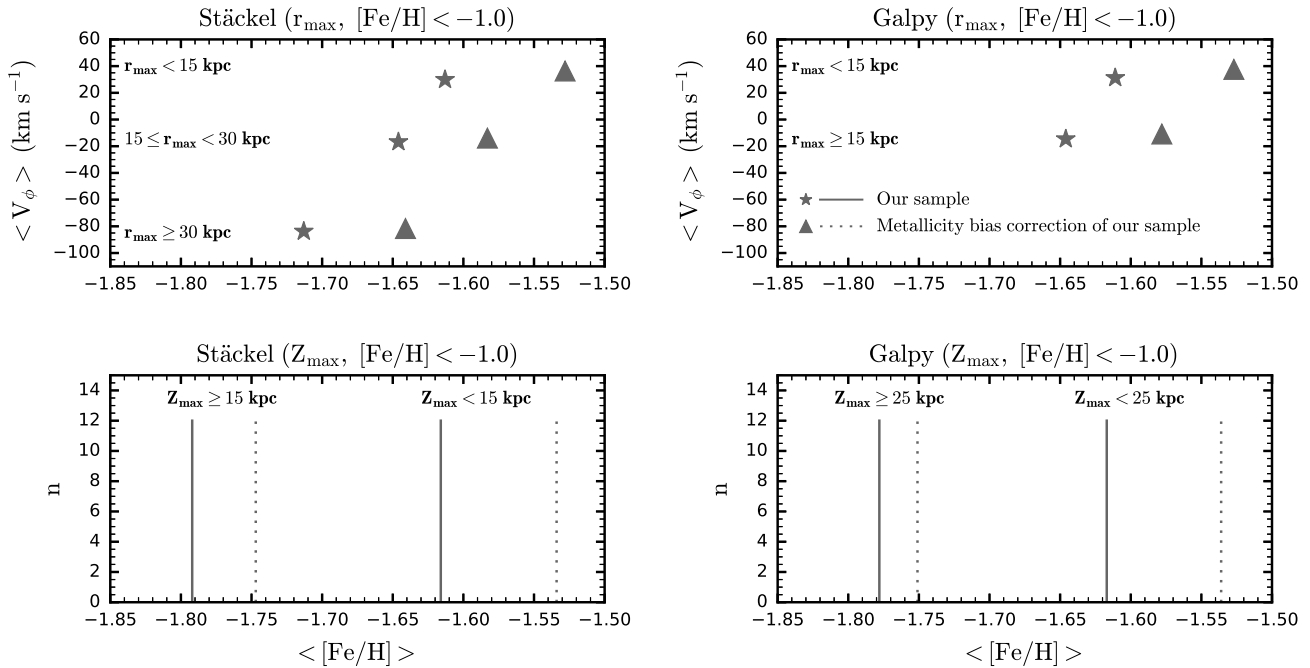}
\caption{Same as in Figure \ref{figure14}, but for the checks on the 
impacts of metallicity bias by selection effect of our sample. In the 
top panels, our sample and the one corrected for metallicity bias by selection 
effect are displayed with a gray star symbol and triangle, respectively. 
In the bottom panels, our sample is indicated by a gray solid line, while the one corrected 
for metallicity bias by selection effect is denoted by a gray dotted line.}
\label{figure16}
\end{figure*}

After construction of an unbiased MDF for our sample, we obtained 
the  mean \vphi\ and [Fe/H] from the 100 simulated samples in the same \rmax\ 
and \zmax\ regions as in Figure \ref{figure14}, as shown 
in Figure \ref{figure16}, for both the \stackel\ and Galpy 
potentials. The top panels of Figure \ref{figure16} display the mean 
\vphi\ versus mean [Fe/H] for stars with [Fe/H] $<$ --1.0 in the regions of \rmax\ $<$ 15 kpc, 
15 $\leq$ \rmax\ $<$ 30 kpc, and \rmax\ $\geq$ 30 kpc for the \stackel\ potential (left), 
and \rmax\ $<$ 15 kpc, \rmax\ $\geq$ 15 kpc for the Galpy potential (right). 
The gray star symbol indicates our original sample, while the triangle
denotes the sample corrected for the target selection
effect. From inspection of the figure, it is clear there is almost no difference in 
\vphi\ between our original sample and the unbiased sample for both potentials. Even though 
there is a shift to higher metallicity, the magnitude of the shift is less than 0.1 
dex, which is smaller than the uncertainty of the metallicity of individual stars. 
We also note that this small metallicity shift is not correlated with \rmax.

The bottom panels represent the distribution of mean [Fe/H] for stars with 
[Fe/H] $<$ --1.0 in the ranges of \zmax\ $\geq$ 15 kpc and \zmax\ $<$ 15
kpc for the \stackel\ potential (left), and \zmax\ $\geq$ 25
kpc and \zmax\ $<$ 25 kpc for the Galpy potential (right).
The gray solid line is for our original sample and the dotted line is
for the unbiased sample. The mean metallicity is shifted to slightly
higher values, slightly less so for high \zmax\ in both potentials. Once
again, however, the scale of the shift is less than 0.1 dex. We
conclude from this exercise that our sample of stars, even after being
corrected for any target selection effect, exhibits clear retrograde
motion for \rmax\ $\geq$ 30 kpc, as well as the transition in metallicity
in both \rmax\ and \zmax\ that we identified in our original sample.

\section{Discussion} \label{sec:discussion}

Traditionally, the separation of distinct stellar components in the
solar neighborhood has been carried out kinematically, by consideration
of the velocity components for individual stars in a Toomre diagram
(e.g., \citealt{venn2004, bonaca2017}). In our study, in order to
characterize different stellar components in a local sample of stars, we
scrutinized the distribution of rotation velocity as a function of
\rmax, which is more correlated with the total energy of stars than the
orbital parameter \zmax. 

Through this investigation, we found that the retrograde motions become
as large as \vphi\ = --150 \kms\ at the largest \rmax, and that the
counter-rotating signature varies more strongly with \rmax\ than with
\zmax. These results are consistent with the work of \citet{kafle2017},
who show larger retrograde motion at greater Galactocentric distances,
11 $< r <$ 15 kpc, based on an in situ sample of main-sequence turn-off stars.

We also identified the transition of one stellar component to another,
occurring at \rmax\ $\sim$ 30 kpc, in the \stackel\ potential. However,
because we did not find such a transition in the Galpy potential, we
decided to separate the halo stars in our sample into the IHP and OHP
at \rmax\ = 30 kpc, as derived from the \stackel\ potential. 

Within the OHP, the more metal-poor stars ([Fe/H] $<$ --1.7) exhibit
more retrograde motions, independent of \zmax. These stars have
characteristics of more polar motions at higher \zmax, which means 
that they have more tangential motions for high-inclination 
orbits while possessing more radial motions at lower \zmax. On the other hand, 
the more metal-rich stars (--1.7 $\leq$ [Fe/H] $<$ --1.0) are located over 
all \zmax, with both radial and retrograde motions, as can be seen in the 
right-most panel of Figure \ref{figure8}.

The derived chemical and kinematic features of our sample can be understood 
as follows, in terms of their assembly over Galactic history.
Satellite galaxies are affected by both dynamical friction and tidal
forces when interacting with the MW, and their stars are stripped off, losing
energy and angular momentum. The kinematic properties of stripped
stars are considered to be equal to those of their parent galaxies when
they are fully accreted into the MW (e.g., \citealt{quinn1986,
vandenbosch1999, ruchti2014, amorisco2017}). Due to weak dynamical
friction and self-gravity, an accreted low-mass satellite galaxy loses
its stars in outer region of the MW by tidal stripping, whereas a
high-mass satellite galaxy sinks farther into the inner region of the MW
due to the stronger influence of dynamical friction on its more
numerous stars, prior to tidal disruption (\citealt{amorisco2017}). 
In the case of a massive satellite galaxy, with mass on the
order of 10\% of the MW, the satellite galaxy with a prograde motion contributes more stars to 
the interior of the MW due to stronger dynamical friction than a
satellite galaxy with a retrograde motion (\citealt{quinn1986}). Results similar
to those above have been found in the series of papers from
\citet{tissera2018}, and references therein. 

The stars stripped off from a lower-mass satellite galaxy with a large orbital 
inclination and high tangential velocity escape at larger distance from
the Galactic plane than those of a high-mass satellite galaxy,
regardless of its orbital direction. By comparison, the stars stripped
from a satellite galaxy with high radial velocity have small \zmax\
regardless of its orbital inclination. Consequently, for local stars,
the stripped stars from a low-mass satellite galaxy with a retrograde
motion are expected to exhibit retrograde motions at the highest (least
bound) energies, whereas those from a high-mass satellite galaxy with a
prograde motion are expected to move with prograde motions in the inner
region of the MW. As a result, if the stars of a low-mass satellite
galaxy are more metal-poor than those of a high-mass satellite galaxy
(as they are expected to be, due to the difference in their star-formation
histories), one could explain the observed relationship between
metallicity and rotational velocity as a function of \rmax. In addition,
as metal-rich stars of our sample have high radial velocities with small
\zmax, the transition of metallicity may be more distinct with \zmax\
than \rmax.

Observationally, our results agree with those inferred from a number of other
recent studies. For example, using main-sequence and blue
horizontal-branch stars at heliocentric distances $\leq$ 10 kpc in the
SDSS-$Gaia$ DR1 catalog, \citet{myeong2018} show that the more metal-rich
stars ([Fe/H] $>$ --1.9) are well-populated in the high-energy (large
\rmax) and retrograde regions. \citet{myeong2019} identify 
these stars as remnant of the Sequoia Event discernible in the retrograde 
stellar substructures. The metal-poor stars, however, are more
sparsely populated in these regions, in contrast to our result (see the
green histograms of the right panel of Figure \ref{figure8}). Similarly,
by analyzing main-sequence stars from the same catalog,
\citet{belokurov2018} find that the metal-rich ([Fe/H] $>$ --1.7) stars
exhibit strong overdensities in the region of high $V_{\rm r}$
velocity, through a multi-Gaussian decomposition of their velocity
distributions. This is consistent with our results, as seen in the $V_{\rm
r}$ velocity distribution (shown in orange in Figure \ref{figure8}).
Furthermore, the majority of stars associated with
$Gaia$-Enceladus, as described by \citet{helmi2018}, are also distributed
in the region of 15 $<$ \rmax\ $<$ 30 kpc and --200 $<$
\vphi\ $<$ 10 \kms, which is a similar to our results. In
addition, many stars with high radial velocity (\rmax\ ${\geq}$ 30 kpc) in the study of \citet{belokurov2018}
are considered to be stripped from
$Gaia$-Enceladus at high orbital energies, while 
counter-rotational stars with \vphi\ $<$ --200 \kms\ are from the Sequoia.

Additionally, we notice the enhancement of stars with \vphi\ = 50 -- 200
\kms\ in the region of \zmax\ $<$ 5 kpc of the IHP in Figure
\ref{figure8}. This may be due to a metal-weak thick-disk (MWTD)
population. \citet{carollo2010} identify such a population, which they
speculate could be kinematically independent of the canonical thick
disk, with net rotation \vphi\ = 100 -- 150 \kms\ within 5 kpc of the
Galactic plane using the SDSS DR7, which has characteristics similar to
those of the enhancement found in our sample. Additional data from
\citet{beers2014} and \citet{carollo2019} bolster this interpretation.
Another explanation for the enhancement may be in situ stars formed in a
protodisk and heated up as described by \citet{mccarthy2012}. Using the
sample of 412 MW-mass disk galaxies in the Galaxies-Intergalactic Medium
Interaction Calculation suite of cosmological hydrodynamical
simulations, they show that in situ stars with prograde motions dominate
at $r$ $\leq$ 30 kpc, whereas the outer halo is dominated by accreted
stars. They obtain a median rotational lag of --0.35 (normalized by
$V_{\rm LSR}$) of the in situ component in the stacked sample limited to
the solar neighborhood. Adopting $V_{\rm LSR}$ = 220 \kms\ for the MW,
the median rotational velocity of the in situ stars is 143 (= $220
\times (1-0.35)$) \kms, which is in the range we find as well.

Furthermore, our sample of stars suggests that, due to the different
orbital characteristics between metal-poor and metal-rich stars, the
metal-poor stars are dominant beyond \zmax\ = 15 kpc in the inner halo
and above \zmax\ = 25 kpc in the outer halo, when separating the halo
regions at \rmax\ = 30 kpc. It is also possible to identify the shift of
the MDF to be more metal-poor for \rmax\ $>$ 30 kpc. As a result, the
OHP is more metal-poor than the IHP. These characteristics agree with
the features found in the in situ halo sample of red giants by
\citet{chen2014}, who report that the transition from the inner to outer
halo occurs at a vertical distance \z\ $\sim$ 20 kpc from the Galactic
plane and at a distance of $r \sim$ 35 kpc from the Galactic center.
Although the transition regions differ, our findings on the dual nature
of the Galactic halo are also supported by the analyses of photometric
samples, which suffer less from potential target-selection bias. Those
results indicate that the halo out to $\sim$ 10 kpc from the Sun is
comprised of two stellar components (\citealt{an2013, an2015}), and the
metallicity of the halo varies between 10 and 20 kpc in Galactocentric
distance (\citealt{dejong2010}).

\section{Summary} \label{sec:summary}

We have presented a study of the dependence of derived Galactic halo
properties, employing two well-known, frequently used Galactic potentials (the
\stackel\ and the Galpy), based a local sample of stars
selected from the spectrophotometric calibration and telluric standard stars of SDSS
DR12, using the criteria of \citet{carollo2010}. 

We found that the shape of the MDF and the rotational velocity distribution
abruptly change at 15 kpc of \zmax\ in the \stackel\ potential, which
suggests that the transition from the inner to outer halo occurs at that
distance. We further confirmed that the stars in the outer-halo region
show a retrograde motion of \vphi\ = --60 \kms\ and a mean metallicity
of [Fe/H] = --1.9. In contrast, when the Galpy potential is used, even though we
identified the transition of the metallicity distribution at \zmax\ = 25
kpc, there is no noticeable retrograde motion found. It is interesting
to note that the unbound stars under the Galpy potential show retrograde
motions regardless of \zmax, whereas the bound and unbound stars in the
same \zmax\ bins exhibit different MDFs. At any rate, as these
discrepancies arise from the different energies of individual stars, we
explored halo properties with \rmax, which is more correlated with
orbital energy.

The exploration of our sample with \rmax\ revealed that
the MDF and rotational velocity distribution are shifted to more
metal-poor and retrograde motions at \rmax\ = 30 kpc in the \stackel\
potential. According to these results, we have newly defined the OHP as 
stars with \rmax\ $>$ 30 kpc. By this division, we found that the OHP 
consists of a mixture of metal-rich and metal-poor stars, 
both of which are found among the stars with retrograde motions. 
Furthermore, the two stellar components exhibit different kinematic
characteristics, depending on \zmax, in the sense that the metal-rich
stars exhibit large radial motions regardless of \zmax, whereas the
metal-poor stars have primarily radial motions at lower
\zmax, but more polar motions at higher \zmax. We also found that the 
profiles of the rotation velocity and metallicity over \rmax\ for the
OHP exhibit a retrograde motion as low as \vphi\ = --150 \kms\ and
metallicity down to [Fe/H]= --1.7. Conversely, the halo stars under the
Galpy potential do not exhibit such a dual nature in either the MDFs or
rotational velocity distribution.

The reason for the dissimilar characteristics inferred for halo stars
between the \stackel\ and the Galpy potential analyses stems from the
fact that stars with high retrograde motions in the \stackel\ potential
are unbound in the Galpy potential, and stars with low rotational
velocities reach to more distant \zmax\ and \rmax, due to the shallower
Galpy potential. Because the observed nature of the halo differs, depending
on the adopted Galactic potential, it is clear that a more realistic
Galactic potential is required going forward, particularly when using
only relatively local samples of halo stars.

\acknowledgments

We thank an anonymous referee for a careful review of this paper,
which has improved the clarity of its presentation.

Funding for SDSS-III was provided by the Alfred P. Sloan Foundation, the
Participating Institutions, the National Science Foundation, and the U.S.
Department of Energy Office of Science. The SDSS-III Web site is
http://www.sdss3.org/.

Y.S.L. acknowledges support from the National Research Foundation (NRF) of
Korea grant funded by the Ministry of Science and ICT (No.2017R1A5A1070354
and NRF-2018R1A2B6003961). T.C.B. acknowledges partial support for this work
from grant PHY 14-30152; Physics Frontier Center/JINA Center for the Evolution
of the Elements (JINA-CEE), awarded by the U.S. National Science
Foundation.

\end{document}